\input epsf
\documentstyle{amsppt}
\newcount\mgnf\newcount\tipi\newcount\tipoformule\newcount\greco 
\tipi=2          
\tipoformule=0   

\global\newcount\numsec\global\newcount\numfor
\global\newcount\numapp\global\newcount\numcap
\global\newcount\numfig\global\newcount\numpag
\global\newcount\numnf
\global\newcount\numtheo

\def\SIA #1,#2,#3 {\senondefinito{#1#2}%
\expandafter\xdef\csname #1#2\endcsname{#3}\else
\write16{???? ma #1,#2 e' gia' stato definito !!!!} \fi}

\def \FU(#1)#2{\SIA fu,#1,#2 }

\def\etichetta(#1){(\veroparagrafo.\veraformula)%
\SIA e,#1,(\veroparagrafo.\veraformula) %
\global\advance\numfor by 1%
\write15{\string\FU (#1){\equ(#1)}}%
\write16{ EQ #1 ==> \equ(#1)  }}

\def\etichettat(#1){\veroparagrafo.\veratheorema:%
\SIA e,#1,{\veroparagrafo.\veratheorema} %
\global\advance\numtheo by 1%
\write15{\string\FU (#1){\thu(#1)}}%
\write16{ TH #1 ==> \thu(#1)  }}

\def\etichettaa(#1){(A\veraappendice.\veraformula)
 \SIA e,#1,(A\veraappendice.\veraformula)
 \global\advance\numfor by 1
 \write15{\string\FU (#1){\equ(#1)}}
 \write16{ EQ #1 ==> \equ(#1) }}
\def\getichetta(#1){Fig. \verafigura
 \SIA g,#1,{\verafigura}
 \global\advance\numfig by 1
 \write15{\string\FU (#1){\graf(#1)}}
 \write16{ Fig. #1 ==> \graf(#1) }}
\def\retichetta(#1){\numpag=\pgn\SIA r,#1,{\verapagina}
 \write15{\string\FU (#1){\rif(#1)}}
 \write16{\rif(#1) ha simbolo  #1  }}
\def\etichettan(#1){(n\verocapitolo.\veranformula)
 \SIA e,#1,(n\verocapitolo.\veranformula)
 \global\advance\numnf by 1
\write16{\equ(#1) <= #1  }}

\newdimen\gwidth
\gdef\profonditastruttura{\dp\strutbox}
\def\senondefinito#1{\expandafter\ifx\csname#1\endcsname\relax}
\def\BOZZA{
\def\alato(##1){
 {\vtop to \profonditastruttura{\baselineskip
 \profonditastruttura\vss
 \rlap{\kern-\hsize\kern-1.2truecm{$\scriptstyle##1$}}}}}
\def\galato(##1){ \gwidth=\hsize \divide\gwidth by 2
 {\vtop to \profonditastruttura{\baselineskip
 \profonditastruttura\vss
 \rlap{\kern-\gwidth\kern-1.2truecm{$\scriptstyle##1$}}}}}
\def\verapagina{
{\romannumeral\number\numcap}.\number\numsec.\number\numpag}}

\def\alato(#1){}
\def\galato(#1){}
\def\veroparagrafo{\number\numsec}\def\veraformula{\number\numfor}
\def\veraappendice{\number\numapp}
\def\verapagina{\number\pageno}\def\veranformula{\number\numnf}
\def\verafigura{{\romannumeral\number\numcap}.\number\numfig}
\def\verocapitolo{\number\numcap}\def\veranformula{\number\numnf}
\def\veratheorema{\number\numtheo}
\def\Eqn(#1){\eqno{\etichettan(#1)\alato(#1)}}
\def\eqn(#1){\etichettan(#1)\alato(#1)}
\def\TH(#1){{\etichettat(#1)\alato(#1)}}
\def\thv(#1){\senondefinito{fu#1}$\clubsuit$#1\else\csname fu#1\endcsname\fi} 
\def\thu(#1){\senondefinito{e#1}\thv(#1)\else\csname e#1\endcsname\fi}
\def\ver{\veroparagrafo}
\def\Eq(#1){\eqno{\etichetta(#1)\alato(#1)}}
\def\eq(#1){\etichetta(#1)\alato(#1)}
\def\Eqa(#1){\eqno{\etichettaa(#1)\alato(#1)}}
\def\eqa(#1){\etichettaa(#1)\alato(#1)}
\def\dgraf(#1){\getichetta(#1)\galato(#1)}
\def\drif(#1){\retichetta(#1)}

\def\eqv(#1){\senondefinito{fu#1}$\clubsuit$#1\else\csname fu#1\endcsname\fi}
\def\equ(#1){\senondefinito{e#1}\eqv(#1)\else\csname e#1\endcsname\fi}
\def\graf(#1){\senondefinito{g#1}\eqv(#1)\else\csname g#1\endcsname\fi}
\def\rif(#1){\senondefinito{r#1}\eqv(#1)\else\csname r#1\endcsname\fi}
\def\bib[#1]{[#1]\numpag=\pgn
\write13{\string[#1],\verapagina}}

\def\include#1{
\openin13=#1.aux \ifeof13 \relax \else
\input #1.aux \closein13 \fi}

\openin14=\jobname.aux \ifeof14 \relax \else
\input \jobname.aux \closein14 \fi
\openout15=\jobname.aux
\openout13=\jobname.bib


\ifnum\tipoformule=1\let\Eq=\eqno\def\eq{}\let\Eqa=\eqno\def\eqa{}
\def\equ{}\fi


{\count255=\time\divide\count255 by 60 \xdef\hourmin{\number\count255}
        \multiply\count255 by-60\advance\count255 by\time
   \xdef\hourmin{\hourmin:\ifnum\count255<10 0\fi\the\count255}}

\def\oramin{\hourmin }

\def\data{\number\day/\ifcase\month\or january \or february \or march \or
april \or may \or june \or july \or august \or september
\or october \or november \or december \fi/\number\year;\ \oramin}

\def\titdate{ \ifcase\month\or January \or February \or March \or
April \or May \or June \or July \or August \or September
\or October \or November \or December \fi \number\day, \number\year;\ \oramin}

\setbox200\hbox{$\scriptscriptstyle \data $}

\newcount\pgn \pgn=1
\def\foglio{\number\numsec:\number\pgn
\global\advance\pgn by 1}
\def\foglioa{A\number\numsec:\number\pgn
\global\advance\pgn by 1}

\footline={\rlap{\hbox{\copy200}}\hss\tenrm\folio\hss}


\global\newcount\numpunt

\magnification=\magstephalf
\baselineskip=16pt
\parskip=8pt

\voffset=2.5truepc
\hoffset=0.5truepc
\hsize=6.1truein
\vsize=8.4truein 
\def\rightheadline{\it  {tralala}\hfil\tenrm\folio}
\def\leftheadline{\tenrm \folio \hfil\it  {Section $\ver$}}

\def\a{\alpha}
\def\b{\beta}
\def\d{\delta}
\def\e{\epsilon}

\def\f{\phi}
\def\g{\gamma}

\def\l{\lambda}

\def\s{\sigma}

\def\o{\omega}
\def\D{\Delta}

\def\G{\Gamma}
\def\O{\Omega}

\def\1{{1\kern-.25em\roman{I}}}
\def\eu{{1\kern-.25em\roman{I}}}
\def\f1{{1\kern-.25em\roman{I}}}

\def\R{{\Bbb R}}  
\def\N{{\Bbb N}}  
\def\P{{\Bbb P}}  
\def\Z{{\Bbb Z}}  
\def\Q{{\Bbb Q}}  
\def\E{{\Bbb E}}  

\def\dist{\,\roman{dist}}

\let\cal=\Cal
\def\AA{{\cal A}}
\def\BB{{\cal B}}

\def\DD{{\cal D}}
\def\EE{{\cal E}}

\def\KK{{\cal K}}

\def\NN{{\cal N}}

\def\PP{{\cal P}}

\def\RR{{\cal R}}
\def\SS{{\cal S}}

\def\WW{{\cal W}}

\def\chap #1#2{\line{\ch #1\hfill}\numsec=#2\numfor=1\numtheo=1}

\def\un #1{\underline{#1}}

\def\ba{{\backslash}}

\def\wt{\widetilde}
\def\wh{\widehat}

\def\limlaw{\buildrel \DD\over\rightarrow}


\def\note#1{\footnote{#1}}

\def\frac#1#2{{#1\over #2}}

\def\text#1{\quad{\hbox{#1}}\quad}
\def\newpage{\vfill\eject}
\def\proposition #1{\noindent{\thbf Proposition #1}}

\def\theo #1{\noindent{\thbf Theorem {#1} }}

\def\lemma #1{\noindent{\thbf Lemma {#1} }}

\def\proof{{\noindent\pr Proof: }}

\def\endproof{$\diamondsuit$}
\def\remark{\noindent{\bf Remark: }}
\def\thanks{\noindent{\bf Acknowledgements: }}

\font\pr=cmbxsl10

\font\thbf=cmbxsl10 scaled\magstephalf

\font\ch=cmbx12
\font\ftn=cmr8

\font\it=cmti10
\font\bf=cmbx10

\font\refer=cmr9
\newfam\msafam
\newfam\msbfam
\newfam\eufmfam
%
%
%
\def\hexnumber#1{%
  \ifcase#1 0\or 1\or 2\or 3\or 4\or 5\or 6\or 7\or 8\or
  9\or A\or B\or C\or D\or E\or F\fi}
\font\tenmsa=msam10
\font\sevenmsa=msam7
\font\fivemsa=msam5
\textfont\msafam=\tenmsa
\scriptfont\msafam=\sevenmsa
\scriptscriptfont\msafam=\fivemsa        
\edef\msafamhexnumber{\hexnumber\msafam}%
%
%
\mathchardef\restriction"1\msafamhexnumber16
\mathchardef\ssim"0218
\mathchardef\square"0\msafamhexnumber03
\mathchardef\eqd"3\msafamhexnumber2C
\def\QED{\ifhmode\unskip\nobreak\fi\quad
  \ifmmode\square\else$\square$\fi}            
\font\tenmsb=msbm10
\font\sevenmsb=msbm7
\font\fivemsb=msbm5
\textfont\msbfam=\tenmsb
\scriptfont\msbfam=\sevenmsb
\scriptscriptfont\msbfam=\fivemsb
\def\Bbb#1{\fam\msbfam\relax#1}    
\font\teneufm=eufm10
\font\seveneufm=eufm7
\font\fiveeufm=eufm5
\textfont\eufmfam=\teneufm
\scriptfont\eufmfam=\seveneufm
\scriptscriptfont\eufmfam=\fiveeufm

\def\({\left(}
\def\){\right)}
%
%
%

\let\Z=\integer

%

\font\tit=cmbx12
\font\aut=cmbx12
\font\aff=cmsl12
\def\s{\char'31}
\centerline{\tit  Rigorous results on some simple spin glass models }
\vskip.2truecm
\centerline{\tit}
\vskip.2truecm 
\vskip1.5truecm
\centerline{\aut Anton Bovier 
\note{ e-mail:
bovier\@wias-berlin.de} \note{Research supported in part by the DFG in
the Concentration program ``Interacting stochastic systems of high 
complexity''.}
}
\vskip.1truecm
\centerline{\aff Weierstra\s {}--Institut}
\centerline{\aff f\"ur Angewandte Analysis und Stochastik}
\centerline{\aff Mohrenstrasse 39, D-10117 Berlin, Germany}
\vskip.4truecm
\centerline{\aut  Irina Kurkova\note{\ftn
e-mail: kourkova\@ccr.jussieu.fr}}
\vskip.1truecm
\centerline{\aff Laboratoire de Probabilit\'es et Mod\`eles
Al\'eatoires}
\centerline{\aff Universit\'e Paris 6}
\centerline{\aff 4, place Jussieu, B.C. 188} 
\centerline{\aff 75252 Paris, Cedex 5, France}

\vskip1truecm\rm
\def\s{\sigma}
\noindent {\bf Abstract:} In this paper we review some recent rigorous 
results that provide an essentially complete solution of a class of
spin glass models introduced by Derrida in the 1980ies. These models
are based on Gaussian random processes on $\{-1,1\}^N$ whose
covariance is a function of a ultrametric distance on that set. We
prove the convergence of the free energy as well as the Gibbs measures
in an appropriate sense. These results confirm fully the predictions
of the replica method including in 
 situations where continuous replica symmetry
breaking takes place.


 \vfill
$ {} $

\newpage

{\headline={\ifodd\pageno\rightheadline \else \leftheadline \fi}}
\def\rightheadline{\it  {Simple spin glasses}\hfil\tenrm\folio}
\def\leftheadline{\tenrm \folio \hfil\it  {Section \ver}}

\def\cov{\hbox{\rm cov}\,} 
\chap{1. Introduction.}1

In spite of considerable recent progress \cite{T1,T3,T4,GT,G}, 
there remains a considerable gap between the heuristic understanding of 
mean field spin glasses such as the Sherrington-Kirkpatrick model \cite{SK}
 (see \cite{MPV}) ,
and the mathematical understanding of the properties of such models. 
We have now a reasonably good insight in situations when the 
so-called ``replica symmetric'' solution is expected to hold, 
but already solution of a model with one-step replica symmetry breaking 
has required an enormous effort \cite{T6}. Understanding situations with full
continuous replica symmetry breaking in the context of SK models appears 
presently quite hopeless, even though Guerra \cite{G} has proven very recently 
an extremely interesting result that shows that in the standard SK model, 
Parisi's solution provides a lower bound for the free energy.  
 
In this note I will report on progress in understanding the 
emergence of replica symmetry breaking in the context of a class of ``simple''
spin glass models, introduced by 
Derrida in 1980: the {\it random energy model} (REM)\cite{D1,D2}, and
the {\it generalised random energy
model}(GREM)\cite{D3,DG1,DG2,DG3}. The former consisted of
modelling the random energy landscape as simply i.i.d. Gaussian
random variables on the set of spin configurations, $\{-1,1\}^N$.
This model can be seen formerly as the limit of the so-called
$p$-spin SK-models \cite{SK}, when $p$ tends to infinity
\cite{D1}. In spite of its simplicity, this model has proven to be
a rather instructive toy model, and has received considerable
attention in the mathematical community
\cite{DG3,DW,Ei,OP,GMP,Ru,BKL,KP, B}. Of course, in many respects  this model is mathematically
almost trivial, and physically quite unrealistic, as all the
dependence structure that is present in more realistic models like the
SK model, is absent. The GREM was introduced in view of {\it keeping}
dependence, while simplifying it to a {\it hierarchical} structure
to still yield a mathematically more tractable model. In fact, the
GREM can be seen as a class of models that is obtained by equipping
the hypercube $\{-1,1\}^N$ with a tree structure and an associated
ultra-metric distance, and then considering standardized Gaussian
random fields on the hypercube whose correlation function depends only
on this distance. We will call these models {\it ``Derrida's models''}
in contrast to the {\it ``Sherrington/Kirkpatrick (SK) models''}
where the covariance depends on the Hamming distance, respectively the
overlap $R_N(\s,\s')= N^{-1}\sum_{i=1}^N \s_i\s_j$.

In \cite {DG1}, B.~Derrida and E.~Gardner presented a solution of the model
with finitely many hierarchies
in the sense that they computed the free energy in the thermodynamic
limit. A rigorous derivation of this solution (in a somewhat more elegant form)
was later obtained by Cappocaccia et al. \cite{CCP}.
Derrida and Gardner also considered the limit of their formulae when the
number of hierarchies tends to infinity. They argued that for suitable choices
of the covariance function, this limits yield approximations for the
standard $p$-spin SK models, even though, as they point out, the quality of
the approximations is not spectacular.

In this paper we review recent results obtained in \cite{BKL,BK1,BK2,BK3}
that give an essentially complete solution confirming the results of the replica
method for all these models. In Section 2 we present first in detail 
the rather simple case of the REM which will serve as a pedagogical example. 
In Section 3 we then turn to the general class of Derrida's models.

\medskip
\chap{2.  The random energy model.\hfill}2

The random energy model, introduced by Derrida [D1,D2] can be considered
as the ultimate toy model of a disordered system. In 
this model, rather little is left of the structure of interacting
spins, but we will still be able to gain a lot of insight into the 
peculiarities 
of disordered systems by studying this simple system. For  rigorous
work on the REM see e.g. [Ei,OP,GMP,DW,BKL,T5].

The REM is a model with state space $\SS_N=\{-1,+1\}^N$. For fixed $N$, 
the Hamiltonian is 
given by
$$
H_N(\s)=-\sqrt N X_\s
\Eq(C.15)
$$
where $X_\s$, is a family of $2^N$ i.i.d. centered normal 
random variables.
\smallskip
\line{\bf \ver.1. The free energy.\hfill}

Before turning to the question of Gibbs measures, we turn to the
simpler question of analysing in some detail the partition
function. In this model, the partition function is of course just the
sum of i.i.d. random variables, i.e.
$$
Z_{\b,N}\equiv 2^{-N}\sum_{\s\in \SS_N}e^{\b \sqrt N X_\s}
\Eq(C.16)
$$
One usually asks first for the exponential asymptotics of this
quantity, i.e. one introduces the {\it free energy},
$$
F_{\b,N}\equiv -\frac 1N \ln Z_{\b,N}
\Eq(C.17)
$$   
and tries to find its limit  as $N\uparrow \infty$.
Let me mention that in general mean field spin glasses, the existence of 
the limit even of the averaged free energy has been a long standing open 
problem. While writing this note, a preprint by Guerra and Toninelli
\cite{GT}  has 
appeared in which a  simple and clever proof of the existence of the limit
in a rather large class of mean field spin glass models is given.

In this simple model one can  compute this
limit exactly. In fact it was found be Derrida \cite{D1} that: 

\theo {\TH(4.1)}{\it In the REM,
$$
\lim_{N\uparrow\infty}\E F_{\b,N} = \cases -\frac{\b^2}2,&\hbox{\rm for}\,\,\,\b\leq 
\b_c\cr
-\frac{\b_c^2}2-(\b-\b_c)  \b_c ,&\hbox{\rm for}\,\,\,
\b\geq\b_c\endcases
\Eq(C.18.0)
$$
where $\b_c=\sqrt{2\ln 2}$.
}

\smallskip
\line{\bf \ver.2. Fluctuations and limit theorems.\hfill}
 
Knowing the free energy is important, but, as one may expect, it
is not enough to understand the properties of the Gibbs measures
completely. It is the analysis of the fluctuations of the free energy
that will reveal, as we will see, the necessary information.
In the REM this can be done using classical results from the theory of extreme
value statistics. The proofs are, nonetheless, quite cumbersome, and 
 may be found in
in
[BKL] or [B].

\theo {\TH(4.2)}{\it The partition function of the REM has the following 
fluctuations:
\item{(i)} If $\b<\sqrt{\ln 2/2}$, then 
$$
e^{\frac N2(\ln 2-\b^2)}\ln \frac {Z_{\b,N}}{\E Z_{\b,N}} \limlaw
\NN(0,1).
\Eq(4.22)
$$
\item{(ii)} If $\beta=\sqrt{\ln 2/2}$,  then 
 $$ \sqrt{2} 
 e^{\frac N2(\ln 2-\b^2)}\ln \frac {Z_{\b,N}}{\E Z_{\b,N}} \limlaw
\NN(0,1).
\Eq(4.22bis)
$$       
\item{(iii)} Let $\alpha\equiv\beta/{\sqrt{2\ln 2}}$. 
             If $ \sqrt{\ln 2/2}<\b<\sqrt{2\ln 2}$, then
$$
e^{\frac N2(\sqrt{2\ln 2}-\b)^2+\frac\a 2[\ln (N\ln 2)+\ln 4\pi]} \ln 
\frac {Z_{\b,N}}{\E Z_{\b,N}} \limlaw \int_{-\infty}^\infty e^{\a
z}(\PP(dz) - e^{-z}dz),    
\Eq(4.23)
$$
where $\PP$ denotes the Poisson point process\note{For a thorough exposition 
on point processes and their connection to extreme value theory, see in 
particular [Re].} on $\R$ with intensity 
measure $e^{-x}dx$.  
\item{(iv)} If $\beta=\sqrt{2\ln 2}$, then 
$$ 
e^{\frac 12[\ln(N\ln 2)+\ln 4\pi]}\Bigl(
\frac{Z_{\b,N}}{ \E Z_{\b,N}}-\frac{1}{2}+\frac {\ln(N\ln 2)+\ln 4\pi}{4\sqrt{\pi N\ln 2}}
\Bigr)
\!\limlaw\!  \int_{-\infty}^0
e^{ z}(\PP(dz) - e^{-z}dz)   + \int\limits_{0}^\infty e^{ z} \PP(dz). 
\Eq(4.24)
$$  
\item{(v)} If
$\b>\sqrt{2\ln 2}$, then 
$$
e^{-N[\b \sqrt{2\ln 2}-\ln 2]+\frac{\a}2[\ln (N\ln 2)+\ln 4\pi]}
Z_{\b,N}  \limlaw \int\limits_{-\infty}^\infty e^{\a z} \PP(dz)
\Eq(4.25)
$$
and
$$
\ln Z_{\b,N}-\E \ln Z_{\b,N}\limlaw  
\ln \int\limits_{-\infty}^\infty e^{\a z} \PP(dz)
-\E \ln \int\limits_{-\infty}^\infty e^{\a z} \PP(dz).
\Eq(4.25a)
$$
}
\remark Note that  expressions like  $ \int_{-\infty}^0
e^{ z}(\PP(dz) - e^{-z}dz)$ are always understood as
$
\lim_{y\downarrow -\infty}  \int_{y}^0
e^{ z}(\PP(dz) - e^{-z}dz).
$
 All the functionals of the Poisson point process
appearing are almost surely finite random variables. Note that the limit in 
\eqv(4.23) has infinite variance and the one in \eqv(4.25) has infinite mean.

Let us just briefly comment on how these results are obtained. 
In fact, (i) follows from the standard CLT for arrays of independent 
random variables under Lindeberg's condition.

As the Lindeberg condition fails for $2 \b^2\geq \ln 2$, it is
clear
 that we cannot expect a simple CLT beyond this regime. Such a failure of 
a CLT is always a problem related to ``heavy tails'', and results from the
fact
that extremal events begin to influence the fluctuations of the sum. It
appears
therefore reasonable to separate from the sum the terms where $X_\s$ is 
anomalously large. For Gaussian r.v.'s it is well known that the right scale
of separation is given by $u_N(x)$ defined
by
$$
2^N \int\limits_{u_N(x)}^\infty \frac {dz}{\sqrt{2\pi}} e^{-z^2/2} =e^{-x}
\Eq(R.11)
$$
which (for $x>-\ln N/\ln 2$) is equal to (see e.g. [LLR])
$$
u_N(x)= \sqrt{2N\ln 2} +\frac{x}{\sqrt {2N\ln 2}}
-\frac{\ln (N\ln 2)+\ln 4\pi}{2\sqrt{2N\ln 2}}+o(1/\sqrt N),
\Eq(R.12)
$$
$x\in\R$ is a parameter. 
The key to most of what follows relies on the famous result on the
convergence
of the extreme value process to a Poisson point process. 
Let us now introduce the point process on $\R$ given by 
$$
\PP_N \equiv \sum_{\s\in \SS_N} \d_{u_N^{-1}(X_\s)}.
\Eq(R.17)
$$
A classical result from the theory of extreme order statistics (see e.g.
[LLR]) asserts that

\theo{\TH(REM.8)} {\it 
The point process $\PP_N$ converges weakly to a Poisson point process on 
$\R$ with intensity measure $e^{-x}dx$.
}

The key idea is then to split the sum by a cutoff corresponding to whether
$X_\s$ is bigger or smaller than $u_N(x)$; the former can then be represented
as a functional of the extremal process that converges to the Poisson process,
and the latter has to be controlled carefully. The computations are in fact
 quite 
tedious.

If we write
$$
Z_{\beta,N}=Z_{\b,N}^x+(Z_{\b,N}-Z_{\b,N}^x)
\Eq(R.35)
$$
for $\beta \geq \sqrt{2\ln 2}$ 
$$
Z_{\b,N}-Z_{\b,N}^x= 
e^{ N\left[\b \sqrt{2\ln 2}-\ln 2\right]-\frac{\a}2[\ln (N\ln 2)+\ln 4\pi]}
\sum_{\s\in\SS_N}\1_{\{u_N^{-1}(\s)>x\}} e^{\a u_N^{-1}(X_\s)}
\Eq(R.36)
$$
so that for any $x\in \R$,
$$
(Z_{\b,N}-Z_{\b,N}^x)
e^{ -N\left[\b \sqrt{2\ln 2}-\ln 2\right]+\frac{\a}2[\ln (N\ln 2)+\ln
4\pi]}\limlaw \int\limits_{x}^\infty e^{\a z} \PP(dz).
\Eq(R.37)
$$
The remaining term is shown to converge to zero in probability as 
first $N\uparrow\infty$ and then $x\downarrow -\infty$.
\endproof

\bigskip
\smallskip
\line{\bf \ver.3. The Gibbs measure.\hfill}

With our preparation on the fluctuations of the free energy, we have
accumulated enough understanding about the partition function that we
can deal with the Gibbs measures. Clearly, there are a number of ways
of trying to describe the asymptotics of the Gibbs measures. Recalling
the general discussion on random Gibbs measures, it should
be clear that we are seeking a result on the convergence in
distribution of random measures. To be able to state such a results,
we have to introduce a topology on the spin configuration state that makes it
uniformly compact. The usual topology to do this would be 
product topology, and this clearly would be an option here. However,
given what we already know about the partition function, this topology
does not appear suited to give describe the measure appropriately. 
Part (v) of Theorem 2.2 actually implies that 
 the partition function is dominated by a
`few' spin configurations with exceptionally large energy. This is a
feature that should remain visible in a limit theorem. The question we 
therefore must address in mean field models is how to describe a
limiting measure on an infinite dimensional cube that properly reflects the
symmetry (under permutation) of the finite dimensional object, in other words
that views this object in an unbiased way.

A first attempt
consists in mapping the hypercube to the interval $[-1,1]$ via
$$
\SS_N\ni\s\rightarrow r_N(\s)\equiv \sum_{i=1}^N \s_i 2^{-i}\in [-1,1]
\Eq(4.400)
$$
Define the pure point measure $\tilde \mu_{\b,N}$ on $[-1,1]$ by 
$$
\tilde \mu_{\b,N} \equiv \sum_{\s\in\SS_N} \d_{r_N(\s)}\mu_{\b,N}(\s)
\Eq(4.401)
$$
Our results will be expressed in terms of the convergence of these
measures. It will be understood in the sequel that the space of
measures on $[-1,1]$ is  equipped with the topology of weak
convergence, and all convergence results hold with respect to this
topology. 

As the diligent reader will have expected, in the high
temperature phase the limit is the same as for $\b=0$, namely

\theo{\TH(REM.10)} {\it If $\b< \sqrt{2\ln 2}$, then 
$$
\tilde \mu_{\b,N} \rightarrow \frac 12 \l,\text{a.s.}
\Eq(4.402)
$$
where $\l$ denotes the Lebesgue measure on $[-1,1]$.}

\proof Note that we have to prove that for any finite collection of
intervals $I_1,\dots, I_k\subset [-1,1]$, the family of random variables 
$\{\tilde \mu_{\b,N}(I_1),\dots, \tilde \mu_{\b,N}(I_k)\}$ converges
jointly almost surely to $\frac 12|I_1|,\dots, \frac12 |I_k|$. But by
construction these random vectors are independent, so that this will
follow automatically, if we can prove the result in the case
$k=1$. Our strategy is to get first very sharp estimates for a family
of special intervals.

In the sequel we will always assume that $N\geq n$.
We will denote by $\Pi_n$ the canonical projection from $\SS_N$ to $\SS_n$. 
To simplify notation, we will often write $\s_n\equiv \Pi_n\s$ when no 
confusion  can arise. 
For $\s\in\SS_N$, set
$$
a_n(\s)\equiv r_n(\Pi_n\s)
\Eq(C.30)
$$
and 
$$
I_n(\s)\equiv [a_n(\s)-2^{-n},a_n(\s)+2^{-n})
\Eq(C.30.1) 
$$
Note that the union of all these intervals forms a disjoint covering of 
$[-1,1)$. Obviously, these intervals are constructed in such a way that
$$
\tilde\mu_{\b,N} (I_n(\s))=\mu_{\b,N}\left(\{\s'\in\SS_N: \Pi_n(\s')=
\Pi_n(\s)\}\right)
\Eq(C.30.1.1)
$$
The first step in the proof consists in showing that the masses of all
the intervals $I_n(\s)$ are remarkably well approximated by their uniform mass.

\lemma{\TH(REM.11)}{\it  Set $\b'\equiv \sqrt{\frac{N}{N-n}}\b$. 
For any $\s\in\SS_n$,
\item{(i)} If $\b'\leq \sqrt{\frac{\ln 2}2}$, 
$$
|\tilde\mu_{\b,N} (I_n(\s))-2^{-n}|\leq 2^{-n}e^{-(N-n) (\ln 2-{\b'}^2)} Y_{N-n}
\Eq(C.31)
$$
where $Y_{N}$  has bounded variance, as $N\uparrow\infty$.
\item{(ii)}  If $\sqrt{\frac{\ln 2}2}<\b'<\sqrt{2\ln 2}$,
$$
|\tilde\mu_{\b,N} (I_n(\s))-2^{-n}|\leq 2^{-n} e^{-(N-n) (\sqrt{2\ln 2}-
\b')^2/2
-\a \ln (N-n)/2} Y_{N-n}
\Eq(C.32)
$$
where $Y_{N}$ is a random variable with bounded mean modulus. 
\item{(iii)} If $\b=\sqrt {2\ln 2}$, then, for any $n$ fixed,
$$
 |\tilde\mu_{\b,N} (I_n(\s))-2^{-n}|\rightarrow 0 \text{in probability}
\Eq(C.33)
$$
}

\remark
Note that in the sub-critical case, the results imply convergence to
the   uniform product measure on $\SS$  in a {\it very strong
sense}. In particular, the base-size of the cylinders considered
(i.e. $n$) can grow proportionally to $N$, {\it even if
almost sure convergence uniformly for all cylinders is required!} This
is  unusually good. However, one should not be deceived by this fact:
even though seen  from the cylinder masses the Gibbs measures look
like the uniform measure, seen from the point of view of individual
spin configurations  the picture is quite different. In fact, the
measure  concentrates on an {\it exponentially} small fraction of the
full hypercube,  namely those $O(\exp(N(\ln 2-\b^2/2)))$ vertices
that have  energy $\sim \b N$ (Exercise!). 
It is just the fact that this set is still exponentially
large, as long as $\b<\sqrt {2\ln 2}$, and is very uniformly dispersed
over $\SS_N$,  that produces this somewhat paradoxical  effect.  The
rather weak  result in the critical case is not artificial. 
In fact it is not true that almost sure convergence will hold. This
follows e.g. from Theorem 1 in [GMP]. One should of course anticipate
some signature  of the phase transition at the critical point.  

\proof The proof of this lemma is a simple application of the 
first three points in Theorem  \thv(4.2).
Just note that the partial partition functions
$$
Z_{\b,N}(\s_n)\equiv \E_{\s'} e^{\b\sqrt N X_{\s'}}\1_{\Pi_n(\s') =\s_n}
\Eq(C.34)
$$ 
are independent and have the same distribution as $2^{-n} Z_{\b',N-n}$.
But 
$$
\tilde\mu_{\b,N}(I_n(\s_n))=\frac {Z_{\b,N}(\s_n)}{ [Z_{\b,N}-Z_{\b,N}(\s_n)]
+Z_{\b,N}(\s_n)}
\Eq(C.35)
$$
Note that $Z_{\b,N}(\s_n)$ and $[Z_{\b,N}-Z_{\b,N}(\s_n)]$ are independent.
It should now be obvious how to conclude the proof with the help of Theorem 
\thv(4.2).
\endproof

Once we have the excellent approximation of the measure on all of the 
intervals $I_n(\s)$, almost sure convergence of the measure in the weak
topology is  a simple  consequence. Of course, 
this is just a coarse version of the finer results we have, 
and much more precise 
information on the quality of approximation can be inferred from Lemma 
\thv(REM.11). But since the high-temperature phase is not our prime concern,
we will not go further in this direction. 

Somehow much more interesting is the behaviour of the measure at low 
temperatures   that we will discuss now. 
Let us introduce the Poisson point process $\RR$ 
on the strip $[-1,1]\times\R$
with intensity measure $\frac 12 dy\times e^{-x}dx$. If 
$(Y_k,X_k)$ denote the atoms of this process, define a new point process 
$\WW_\a$ on $[-1,1]\times (0,1]$ whose atoms are
$(Y_k, w_k)$, where 
$$
w_k\equiv \frac{e^{\a X_k}}{\int \RR(dy,dx) e^{\a x}}
\Eq(4.406)
$$
for $\a>1$.
Let us note that the process $\wh \WW =\sum_kw_k$ is known in the literature 
as the {\it Poisson-Dirichlet process} with parameter $\a$ \cite{K}.

With this notation we have that

\theo{\TH(REM.12)}{\it If $\b>\sqrt{2\ln 2}$, with $\a=\b/\sqrt {2\ln 2}$,
$$
\tilde \mu_{\b,N} \limlaw \tilde \mu_\b\equiv\int_{[-1,1]\times(0,1]}
\WW_\a(dy,dw) \d_y w
\Eq(4.407)
$$
}

\proof With $u_N(x)$ defined in \eqv(R.12),
we  define the point process $\RR_N$ on $[-1,1]\times\R$ 
by 
$$
\RR_N \equiv \sum_{\s\in\SS_N}\d_{(r_N(\s),u^{-1}_N(X_\s))}
\Eq(4.409)
$$
A standard result of extreme value theory (see [LLR], Theorem 5.7.2)
is easily adapted to yield that
$$
\RR_N\limlaw \RR,\text{as $N\uparrow\infty$}
\Eq(4.4010)
$$
where the convergence is in the sense of weak convergence on the space of
sigma-finite measures endowed with the  (metrizable) topology of vague
convergence. 
Note that 
$$
\mu_{\b,N}(\s)=\frac{e^{\a u_N^{-1}(X_\s)}}{\sum_\s
 e^{\a u_N^{-1}(X_\s)}}=\frac{e^{\a u_N^{-1}(X_\s)}}{\int \RR_N(dy,dx)
e^{\a x}}
\Eq(4.4011)
$$
Since $\int \RR_N(dy,dx)
e^{\a x} <\infty $ a.s., we can define the point process 
$$
\WW_N\equiv \sum_{\s\in\SS_N}\d_{\bigl(r_N(\s), \frac{\exp(\a u_N^{-1}(X_\s))}
{\int \RR_N(dy,dx)
\exp(\a x)}\bigr)}
\Eq(4.4012)
$$
on $[-1,1]\times (0,1]$. Then 
$$
\tilde \mu_{\b,N}=\int \WW_N(dy,dw) \d_y w
\Eq(4.4013)
$$
The only non-trivial point in the convergence proof is to show that the
the contribution to the partition functions in the denominator
from atoms with $u_N(X_\s)<x$ vanishes as $x\downarrow-\infty$. 
But this is precisely what we 
have shown to be the case in the proof of part (v) of Theorem \thv(4.2). 
Standard arguments then imply that first $\WW_N\limlaw\WW$, 
and consequently, \eqv(4.407). \endproof

\remark Note that Theorem \thv(REM.12) contains in particular
the convergence of the Gibbs measure in the product topology on $\SS_N$,
since cylinders correspond to certain subintervals of $[-1,1]$. On the other
hand, it implies that the point process of weights $\sum_{\s\in \SS_N} 
\d_{\mu_{\b,N}(\s)}$ converges in law to the marginal of
$\WW_N$ on $(0,1]$ which is the process introduced by Ruelle \cite{Ru}.
The formulation of Theorem \thv(REM.12) is moreover very much in the spirit of 
the meta-state approach to random Gibbs measures \cite{NS}. 
The limiting measure is a 
measure on a continuous space, and each point measure on this 
set  may appear 
as ``pure state''. The ``meta-state'', i.e. the law of the random 
measure $\tilde\mu_\b$ is  a probability distribution concentrated on the 
countable convex combinations of pure states randomly chosen
by a Poisson point process from an uncountable collection, 
while the coefficients of the convex combination are again random
and selected via another point process.

Let us discuss the properties of the limiting measure $\tilde\mu_\b$.
It is not hard to see that with probability one, the support of $\tilde\mu_\b$
is the entire interval $[-1,1]$. On the other hand, its mass is concentrated 
on a countable set, i.e. the measure is pure point. To see this, consider the
rectangle $A_\e\equiv (\ln \e,\infty)\times [-1,1]$. Clearly, the process $\RR$
restricted to this set has finite total intensity given by 
$\e^{-1}$. i.e. the number total number of atoms in that set is a Poissonian
random variable with parameter $\e^{-1}$. Now if we remove the projection
of these  finitely 
many random points from $[-1,1]$, we will show that the total mass that 
remains goes to zero with $\e$. Clearly, the remaining mass is given by 
$$
\int_{[-1,1]\times(-\infty,\ln \e)}
\RR(dy,dx) \frac{e^{\a x}}{\int \PP(dx')e^{\a x'}}
=\int_{-\infty}^{\ln \e} \PP(dx)  \frac{e^{\a x}}{\int \PP(dx')e^{\a x'}}
\Eq(4.414)
$$
We want to get a lower bound in probability on the denominator. 
The simplest possible bound is obtained by estimating the probability
of the integral by the contribution of the largest atom which of course 
follows the double-exponential distribution.
Thus
$$
\P\left[\int \PP(dx)e^{\a x} \leq Z\right]\leq
 e^{-e^{-\ln Z/\a}}=e^{-Z^{-\frac 1\a}}
\Eq(4.415)
$$
Setting $\O_Z\equiv\{\PP:\int \PP(dx)e^{\a x} \leq Z\}$,
we conclude that, for $\a>1$, 
$$
\eqalign{
\P\left[ \int_{-\infty}^{\ln \e}\PP(dx)  \frac{e^{\a x}}{\int \PP(dx')e^{\a x'}}
>\g\right]
&\leq \P\left[\int_{-\infty}^{\ln \e} \PP(dx)  \frac{e^{\a x}}
{\int \PP(dx')e^{\a x'}}
>\g,\,\O_Z^c\right]+\P[\O_Z]\cr
&\leq \P\left[\int_{-\infty}^{\ln \e} \PP(dx)  e^{\a x}
>\g Z,\,\O_Z^c\right]+\P[\O_Z]\cr
&\leq  \P\left[\int_{-\infty}^{\ln \e} \PP(dx)  e^{\a x}
>\g Z\right]+\P[\O_Z]\cr
&\leq \frac{\E \int_{-\infty}^{\ln \e} \PP(dx)  e^{\a x}}\g +\P[\O_Z]\cr
&\leq \frac {\e^{\a-1}}{(\a-1)\g Z} +e^{-Z^{-\frac 1\a}}
}
\Eq(4.416)
$$
Obviously, for any positive $\g$ it is possible to choose $Z$ as a function
of $\e$ in such a way that the right hand side tends to zero. 
But this implies that with probability one, all of the mass of the measure 
$\tilde \mu_\b$ is carried by a countable set, implying that $\tilde \mu_\b$ 
is pure point. 

So we see that the phase transition in the REM expresses itself via a change 
of the  properties of the infinite volume Gibbs measure mapped to the interval
from Lebesgue measure at high temperatures to a random dense 
pure point measure at low temperatures.

\smallskip
\line{\bf \ver.4. The replica overlap.\hfill}

While the random measure description of the phase transition in the 
REM appears rather nice, one would argue that it ignores fully the 
geometry of the statespace as a hypercube. A neat object to measure look at 
in this respect would be the mass dustribution around a given configuration,
$$
m_\s(t)\equiv \mu_{\b,N}\left(R_N(\s,\s')\geq t\right)
\Eq(mass.1)
$$
where the $\s$ is fixed and the measure $\mu$ refers to the configuration 
$\s'$. $m_\s(\cdot)$ is a probability distribution function on $[-1,1]$. 
As a function of $\s$, this is a measure values random variable. 
Taking the overage of this quantity again with respect to the Gibbs 
distribution of $\s$, we obtain the popular ``overlap distribution'',
$$
f_{\b,N}[\o](dz)\equiv \mu_{\b,N}\left(m_\s(dz)\right)=
 \mu_{\b,N}[\o]\otimes \mu_{\b,N}[\o]\left(R_N(\s,\s')\in dz\right)
\Eq(C.40)
$$
It turns out that a much richer object is obtained by passing to a measure 
valued quantity, namely
$$
\KK_{\b,N}\equiv \sum_{\s\in\SS_N}\mu_{\b,N}(\s)\d_{m_\s(\cdot)}
\Eq(C.41)
$$
This measure  tells
us the probability to see a given miss distribution around oneself, if 
one is distributed with the Gibbs measure. Of course we
have that 
$$
f_{\b,N}[\o](\cdot)=\int \KK_{\b,N}(dm) m(\cdot)
\Eq(mass.2)
$$

Of course, in the REM, one is not likely to see anything very exciting, 
the overlap distribution is asympototically concentrated on the 
values $0$ and $1$ only:

\theo{\TH(REM.15)}{\it 
\item{(i)} For all $\b<\sqrt{2\ln 2}$
$$
\lim_{N\uparrow\infty} f_{\b,N} =\d_0,\text {a.s.}
\Eq(C.41.1)
$$
\item{(ii)} For all $\b>\sqrt{2\ln 2}$
$$
 f_{\b,N} \limlaw  \d_0\left(1- \int \WW(dy,dw) w^2\right)
+ \d_1\int \WW(dy,dw) w^2 
\Eq(C.42)
$$
\item{(iii)} The random measures $\KK_{\b,N}$ converge to
a random probability distribution $\KK_\b$ that is supported on the 
atomic measures with support on $\{0,1\}$, more precisesly
if $\b>\sqrt{2\ln 2}$,
$$
\KK_\b=\int\WW(dy,dw) w\d_{w\d_1+(1-w)\d_0}
\Eq(mass.3)
$$ 
while for $\b< \sqrt{2\ln 2}$, $\KK_\b$ is the Dirac mass on the Dirac mass 
concentrated at $0$.

}

\proof  We will write for any $I\subset [-1,1]$ 
$$
\eqalign{
f_{\b,N}(I) =Z_{\b,N}^{-2} \E_\s\E_{\s'}{\displaystyle \sum_{{t\in
I}\atop{R_N(\s,\s')=t}}e^{\b\sqrt N(X_\s+X_{\s'})}}
}
\Eq(C.43)
$$
First of all, the denominator is bounded from below by $[\wt
Z_{\b,N}(c)]^2$, and, with probability of order
$\d^{-2}\exp(-Ng(c,\b))$, this in turn is larger than 
$(1-\d)^2[\E\wt Z_{\b,N}(c)]^2$.
Now let first $\b<\sqrt{2\ln 2}$. Assume first that $I\subset
(0,1)\cup [-1,0)$. We conclude that 
$$
\eqalign{
\E f_{\b,N}(I) &\leq \frac 1{(1-\d)^2}
\E_\s\E_{\s'} \sum_{{t\in
I}\atop{R_N(\s,\s')=t}}1 + \d^{-2}e^{-g(c,\b)N}
\cr &= \frac 1{\sqrt{2\pi N}}\frac 1{(1-\d)^2} \sum_{t\in
I} \frac{2e^{-N \phi(t)}}{1-t^2}  + \d^{-2}e^{-g(c,\b)N}
}
\Eq(C.44)
$$
for any $\b<c<\sqrt{2\ln 2 }$,
where $\phi:[-1,1]\rightarrow \R$ denotes the Cram\`er entropy function
$$
\phi(t) =\frac {(1+t)}2\ln (1+t)+\frac {(1-t)}2\ln (1-t)
\Eq(C.45)
$$
Here we used of course that, firstly, if $ (1-t) N=2\ell$, $\ell=0,\dots,N$, then
$$
\E_\s\E_{\s'} \1_{R_N(\s,\s')=t}= 2^{-N}{N\choose \ell} 
\Eq(C.45.1)
$$
and, secondly,  Stirling's approximation which implies that 
$$
{N\choose \ell}=\frac 1{\sqrt{2\pi }}\sqrt{\frac N {\ell(N-\ell)}}
\frac{N^N}{\ell^\ell(N-\ell)^{N-\ell}}(1+o(1))
\Eq(Stirl)
$$
valid if $\ell\sim x N$ with $x\in (0,1)$.
 Under our assumptions on $I$, we see immediately from this representation
that the right hand side of \eqv(C.44) is 
clearly
exponentially small in $N$. If $1\in I$, the additional term coming
from $t=1$ gives an exponentially small contribution. This
shows that the measure $f_{\b,N}$ concentrates asymptotically on the
point $0$. This proves \eqv(C.41.1). 

Now let $\b>\sqrt {2\ln 2}$. Here we use the sharper truncations
introduced in \ver.2.  Note first that for any interval $I$ 
$$
\left|f_{\b,N}(I)-Z_{\b,N}^{-2} \E_\s\E_{\s'} \sum_{{t\in
I}\atop{R_N(\s,\s')=t}}\1_{X_\s,X_{\s'}\geq u_N(x)}
e^{\b\sqrt N(X_\s+X_{\s'})}\right|
\leq \frac {2Z_{\b,N}^x}{Z_{\b,N}}
\Eq(C.47)
$$
The proof of Theorem \eqv(4.2) shows that the right hand side of \eqv(C.47) tends zero in
probability
as first $N\uparrow\infty$ and then $x\downarrow -\infty$. 
On the other hand, for $t\neq 1$
$$
\eqalign{
&\P\left[\exists_{\s,\s,:R_N(\s,\s')=t}\,X_\s>u_N(x)\land
X_\s'>u_N(x)\right]
\cr
&\leq \E_\s\1_{{R_N(\s,\s')=t}} \,2^{-2N}\P\left[X_\s>u_N(x)\right]^2= 
\frac 2{\sqrt {2\pi N}\sqrt{1-t^2}} e^{-\phi(t) N}e^{2x}
}
\Eq(C.48)
$$ 
by the definition of $u_N(x)$ (see \eqv(R.11)). This implies again that
any interval $I\subset (0,1)\cup [-1,0)$ will have zero mass. To
conclude the proof it will be enough to compute $f_{\b,N}(1)$. Clearly
$$
f_{\b,N}(1)=\frac {2^{-N}\Z_{2\b,N}}{Z_{\b,N}^2}
\Eq(C.49)
$$ 
By Theorem \thv(4.2), (v), one sees easily that
$$
f_{\b,N}(1)\limlaw \frac {\int e^{2\a z}\PP(dz)}{\left(\int
e^{\a z}\PP(dz)\right)^2}
\Eq(C.50)
$$
Expressing the left hand side of \eqv(C.50) in terms of the point
process $\WW_\a$ defined in \eqv(4.406) yields the expression for the
mass of the atom at $1$; since the only other atom is at zero the full
results follows from the fact that $f_{\b,N}$ is a probability
measure.

The assertions on the measure $\KK_{\b,N}$ are essentially a corollary of the
preceeding results. The fact that $f_\b$ is a sum of $\d_0$ and $\d_1$ 
implies immediately that the probability that $m_\s$ is not such a sum
tends to zero. The explicit formula \eqv(mass.3) is then quite 
straightforward. \endproof

\smallskip
\line{\bf \ver.5. Multi-overlaps and Ghirlanda--Guerra identities. \hfill}

It will be interesting to see that the random measures $\KK_\b$ can be
controlled with the help of some remarkable algebraic identities
that in fact allow us to avoid the detailed analysis of fluctuations
performed in Section \ver.2. 

Let us first note that the convergence of the measures $\KK_{\b,N}$
can be controlled through their moments, which can be written als follows:
$$
\eqalign{
&\E\left(
\int\KK_{\b,N}(dm) m^{k_1}\dots\int\KK_{\b,N}(dm) m^{k_l}\right)
\cr
&=
\E \mu_{\b,N}^{\otimes l}\left(m^{k_1}_{\s^1}(\cdot)\dots
m_{\s^l}^{k_l}(\cdots)\right)
\cr&=\E\mu_{\b,N}^{l+k_1+\dots+k_l}\Bigl(R_N(\s^1,\s^{l+1})\in \cdot,\dots,
R_N(\s^1,\s^{l+k_1})\in \cdot,\dots,\cr
&\quad\quad
\dots,R_N(\s^l,\s^{l+k_1+\dots+k_{l-1}+1})\in \cdot,\dots,
R_N(\s^l,\s^{l+k_1+\dots+k_{l}})\in \cdot\Bigr)
}
\Eq(mass.10)
$$
The right hand side is a (marginal of) the distribution of the
$m(m-1)$ replica overlaps under the averaged product Gibbs measure on 
$m={l+k_1+\dots+k_{l-1}+1}$ independent replicas of the spin variables. 
Thus, if we can show that these multi-replica distributions converge, as 
$N\uparrow\infty$, then the convergence of the measures $\KK_{\b,N}$
will be proven. This is a general fact, which has notheing to do with the 
particular model we look at. In the REM, of course, considerable 
simplification will take place since we know that the 
overlap takes only the values $0$ and one in the limit, and thus instead of
looking at the entore distributions, it will be enough to look at the 
atoms when overlaps equal to $1$. That is to say it will be enough in our case 
to consider the numbers 
$$
\eqalign{
&\E\mu_{\b,N}^{l+k_1+\dots+k_l}\Bigl(R_N(\s^1,\s^{l+1})=1,\dots,
R_N(\s^1,\s^{l+k_1})=1,\dots,\cr
&\quad\quad
\dots,R_N(\s^l,\s^{l+k_1+\dots+k_{l-1}+1})=1,\dots,
R_N(\s^l,\s^{l+k_1+\dots+k_{l}})=1\Bigr)\cr
&=\E\mu_{\b,N}^{l+k_1+\dots+k_l}\Bigl(\s^1=\s^{l+1},\dots,
\s^1=\s^{l+k_1},\dots,\cr
&\quad\quad
\dots,\s^l=\s^{l+k_1+\dots+k_{l-1}+1},\dots,\s^l=\s^{l+k_1+\dots+k_{l}}\Bigr)
\cr&=\E\mu_{\b,N}^{l+k_1+\dots+k_l}\Bigl(\s^1=\s^{l+1}=\dots=\s^{l+k_1},\dots,\cr
&\quad\quad
\dots,\s^l=\s^{l+k_1+\dots+k_{l-1}+1}=\dots=\s^{l+k_1+\dots+k_{l}}\Bigr)
}
\Eq(mass.11)
$$

As we will show now, the multi-overlaps are not independent, but satisfy 
recursion relations that are due to rather 
 general principles. It will be
instructive
to look at them in this simple context. These identities have been
known in the physics literature and a more rigorous analysis is given
in a paper by Ghirlanda and Guerra \cite{GG}. 
Equivalent relations were in fact
derived somewhat earlier by Aizenman and Contucci [AC].
See also \cite{L}.
The importance
of these relations has been underlined by Talagrand \cite{T4,T5}. 
Let us begin
with the simplest instance of these relations.  

\proposition {\TH(GG.1)}{\it For any value of $\b$,
$$
\E\frac{d}{d\b} F_{\b,N}  =-\b(1-\E f_{\b,N}(1))
\Eq(C.51)
$$
}

\proof Obviously,
$$
\E \frac{d}{d\b} F_{\b,N}=-N^{-1}\E \frac {\E_\s \sqrt N X_\s e^{\b \sqrt N
X_\s}} {\E_\s e^{\b \sqrt N
X_\s}} 
\Eq(C.52)
$$
Now if $X$ is standard normal variable, and $g$ any function of at
most polynomial growth, then 
$$
\E [X g(X)]=\E g'(X)
\Eq(C.53)
$$
Using this identity in the right hand side of \eqv(C.52) with respect
to the average over $X_\s$, we get immediately that
$$
\eqalign{
\E \frac {\E_\s \sqrt N X_\s e^{\b \sqrt N
X_\s}} {\E_\s e^{\b \sqrt N
X_\s}}& = N\b\E \left(1-\frac {2^{-N}\E_\s e^{2\b\sqrt NX_\s}}
{(\E_\s e^{\b\sqrt
NX_\s})^2}\right)\cr
&= N\b \E\left(1-\mu_{\b,N}^{\otimes 2}
\left(\1_{\s^1=\s^2}\right)\right)
}
\Eq(C.54)
$$
which is obviously the claim of the lemma.\endproof

In exactly the same way one can prove the following generalisation:

\lemma{\TH(GG.2)}{\it Let $h:\SS_N^n \rightarrow \R$ be any bounded
function of $n$ spins. Then 
$$
\eqalign{
&\frac1{\sqrt N}\E\mu_{\b,N}^{\otimes
n}\left(X_{\s^{k}}h(\s^1,\dots,\s^n)\right)
\cr&=\b\E\mu_{\b,N}^{\otimes
n+1}\left(h(\s^1,\dots,\s^n)\left(\sum_{l=1}^n \1_{\s^k=\s^l}
-n\1_{\s^k=\s^{n+1}}\right)\right)
}\Eq(C.55)
$$
}

\proof Left as an exercise.\endproof

The strength of Lemma \thv(GG.2) comes out when combined with a
factorization result that in turn is a consequence of self-averaging.

\lemma {\TH(GG.3)}{\it Let $h$ be as in the previous lemma.
 For all but possibly a countable number of
values of $\b$,
$$
\lim_{N\uparrow\infty}\frac 1{\sqrt N}\left|
\E\mu_{\b,N}^{\otimes
n}\left(X_{\s^{k}}h(\s^1,\dots,\s^n)\right)-
\E\mu_{\b,N}\left(X_{\s^{k}}\right)
\E\mu_{\b,N}^{\otimes
n}\left(h(\s^1,\dots,\s^n)\right)\right|=0
\Eq(C.56)
$$
}

\proof Let us write 
$$
\eqalign{
&\left(
\E\mu_{\b,N}^{\otimes
n}\left(X_{\s^{k}}h(\s^1,\dots,\s^n)\right)-
\E\mu_{\b,N}\left(X_{\s^{k}}\right)
\E\mu_{\b,N}^{\otimes
n}\left(h(\s^1,\dots,\s^n)\right)\right)^2\cr
&=\left(
\E\mu_{\b,N}^{\otimes
n}\left(\bigl(X_{\s^{k}}-\E\mu_{\b,N}^{\otimes
n}X_{\s^k}\bigr) h(\s^1,\dots,\s^n)\right)\right)^2\cr
&\leq 
\E\mu_{\b,N}^{\otimes n}\left(X_{\s^{k}}-\E\mu_{\b,N}^{\otimes
n}X_{\s^k}\right)^2 \E\mu_{\b,N}^{\otimes n}
\left( h(\s^1,\dots,\s^n)\right)^2
}
\Eq(C.57)
$$
where the last inequality is the Cauchy--Schwarz inequality applied to the 
joint expectation with respect to the Gibbs measure and the disorder.
Obviously the first factor in the last line is equal to 
$$\eqalign{
&\E\left(\mu_{\b,N}(X_\s^2)-[\mu_{\b,N}(X_\s)]^2\right)
+\E\left( \mu_{\b,N}(X_\s)-\E \mu_{\b,N}(X_\s)\right)^2
\cr
& =- \b^{-2}   \E \frac {d^2}{d\b^2} F_{\b,N}
+N
\b^{-2}\E\left( \frac {d}{d\b} F_{\b,N} -\E\frac {d}{d\b} F_{\b,N} \right)^2
}
\Eq(C.58)
$$
We know that $ F_{\b,N}$ converges as $N\uparrow\infty$ and that the 
limit is infinitely differentiable for all $\b\geq 0 $,
 except at $\b=\sqrt{ 2\ln 2}$; moreover, $-F_{\b,N}$ is convex in $\b$. Then 
standard results of convex analysis imply that 
$$
\limsup_{N\uparrow\infty} (- \E \frac {d^2}{d\b^2} F_{\b,N})
=- \frac {d^2}{d\b^2}   \lim_{N\uparrow \infty}\E  F_{\b,N}
\Eq(C.59)
$$
which is finite for all $\b\neq \sqrt{2\ln 2}$. Thus, the first term in 
\eqv(C.58)  will vanish
 when divided by $N$. To see that the coefficient of $N$
of the second term gives a vanishing contribution, we use the general fact that
if the variance of family  of a convex (or concave) functions tends to 
zero, then the same is true for its derivative, except possibly on a countable
 set of values of their  argument. In Theorem \thv(4.2)
 we have more than established
that the variance of $F_{\b,N}$ tends to zero, and hence the result of the 
Lemma is proven. \endproof

If we combine Proposition \thv(GG.1), Lemma \thv(GG.2), and Lemma \thv(GG.3) 
we arrive immediately at 

\proposition {\TH(GG.4)}{\it For all but a countable set of values $\b$,
for any bounded function $h:\SS_N^n\rightarrow\R$,
$$
\eqalign{
\lim_{N\uparrow\infty}
&\Biggl|\E\mu^{\otimes n+1}_{\b,N}\left(h(\s^1,\dots,\s^n)\1_{\s^k=\s^{n+1}}
\right)
\cr&-\frac 1n\E\mu^{\otimes n+1}_{\b,N}\left(h(\s^1,\dots,\s^n)\left(
\sum_{l\neq k}^n \1_{\s^l=\s^k}+\E\mu^{\otimes 2}_{\b,N}(\1_{\s^1=\s^2}
)\right)\right)\Biggr|=0
}
\Eq(C.60)
$$
}

Together with the fact that the product Gibbs measures are 
concentrated only on the sets where the overlaps take values
 $0$ and $1$, \eqv(C.60) permits to compute the distribution of all
higher overlaps in terms of the two-replica overlap. E.g., if we put 
$$
A_n\equiv \lim_{N\uparrow\infty}
\E \mu^{\otimes n}_{\b,N}(\1_{\s^1=\s^2=\dots=\s^n})
\Eq(C.61)
$$
then \eqv(C.60) with $h=\1_{\s^1=\s^2=\dots=\s^n}$
provides the recursion 
$$
\eqalign{
A_{n+1}&=\frac {n-1}n A_n + \frac 1n A_n A_2=A_n\left(1- \frac {1-A_2}n\right) 
\cr&=\prod_{k=2}^n\left(1- \frac {1-A_2}k\right)A_2\cr
&=\frac{\G(n+A_2)}{\G(n+1)\G(A_2)}
}\Eq(C.62)
$$
Note that we can use alternatively Theorem \thv(REM.10) to compute, for 
the non-trivial case $\b>\sqrt{2\ln 2}$,
$$
\lim_{N\uparrow\infty} \mu^{\otimes 2}_{\b,N}(\1_{\s^1=\s^2=\dots=\s^n})
=\int \KK_\b(dm)[m(1)]^{n-1}
\Eq(C.63)
$$
so that \eqv(C.62) implies a formula for the mean of the $n$-th moments of 
$\WW$,
$$
\E\int \WW(dy,dw)w^n=\frac{\G(n+A_2)}{\G(n+1)\G(A_2)}
\Eq(C.64)
$$
where $A_2=\E \int \WW(dy,dw)w^2$. 
This result has been obtained  by a direct computation  by Ruelle
(\cite{Ru}, Corollary  2.2), but its derivation
via the Ghirlanda--Guerra identities shows a way to approach this
problem in a  
different manner that has the potential to give results in more complicated 
situations.\note{ More generally, one may dervive recursion formulas for 
more general moments of Ruelle's process that show that the identies 
\eqv(C.60) determine completely the process of Ruelle in terms of the 
two-overlap distribution function.} 

\bigskip

\chap{3. The Derrida models.}3

The reader of the previous chapter may think that that was 
`much ado about nothing'. First, it was all about independent random variables,
second, we used heavy tools to describe structure that is in fact 
very simple. We will now move towards a class of models that have been 
introduced 17 years ago by Derrida as ``simplified'' spin glass models.
It turns out that while  these models
exhibit structure that is as complex as (and in fact almost identical to ) in 
the Sherrington-Kirkpatrick type spin glasses, they 
 can now be analysed with full rigor whith the help of the tools I 
have explained
in the previous section. The results of these Section cover recent work 
with Irina Kurkova \cite{BK1,BK2,BK3}. 
The purpose of this section is to explain 
how the remarkable universal structures predicted by Parisi's replica 
symmetry breaking scheme arise as a limiting object in a spin glass model.
For further analysis of the limting object itself we refer to  papers by Bolthausen and Sznitman \cite{BoSz} and Bertoin and LeGall \cite{BeLe}.

\medskip
\line{\bf \ver.1. Definitions and basics.\hfill}

As we have already pointed out in the introduction, from a mathematial point of
view it is natural to embed the SK models in the general setting of 
models based on Gaussian processes on the hypercubes $\SS_N$. The special
feature of the SK models in  that context is then that their covariance
depends only on the ``overlap'', $R_N(\s,\s')=\frac 1N (\s,\s')$.

Derrida introduced annother class of 
models that he called {\it Generalized Random Energy models} (GREM)
that can be constructed in full analogy to the SK class by introducing 
annother function charcterizing distance that is to replace the overlap
$R_N$, namely
$$
\dist(\s,\s')\equiv \frac 1N\left(\min(i:\s_i\neq \s'_i)-1\right)
\Eq(cont.0)
$$
To be precise, $\dist$ is an {\it ultrametric} valuation on the set $\SS_N$. 
An ultrametric distance would be given e.g. by a function $D(\s,\s')=\exp(-
\dist(\s,\s'))$.
We will now consider centered Gaussian processes $X_\s$ on $\SS_N$ those 
covariance is given as
$$
\cov(X_\s,X_{\s'})=\E X_\s X_{\s'} =A(\dist(\s,\s'))
\Eq(cont.1)
$$
where $A$ is a probability distribution function on the interval $[0,1]$.

In fact, the original models of Derrida correspond to the special case when
$A$ is the distribution function of a random variable that takes only finitely 
many values, i.e. when $A$ is a monotone increasing step function with 
finitely many steps. However, Derrida
also considered limits when the number of these steps tend to infinty.

The choice of the distance $\dist$ has a number of remarkable effect that 
help to make these models truely solvable. In particular, it allows to 
introduce a continuous time martingale $X_\s(t)$ those marginal at $t=1$
coincides with $X_\s$. This process is simply a Gaussian process on $
\SS_N\times [0,1]$ with covariance
$$
\cov(X_\s(t),X_{\s'}(t'))=t\wedge t'\wedge
A(\dist(\s,\s'))
\Eq(cont.2)
$$
In particular, this gives rise to the integral representation of 
$X_\s$ as
$$
X_\s =\int_0^1 dX_\s(t)
\Eq(cont.3)
$$
where the increments satisfy
$$\E dX_\s(t)d X_{\s'}(t')=dtdt' \d(t-t')\1_{A(\dist(\s,\s'))>t}
\Eq(cont.4)
$$
If $A$ is a step function, this gives rise to a representation in the form
 $$
X_\s\equiv \sqrt{a_1} X_{\s_1}+\sqrt{a_2}X_{\s_1\s_2}+
  \cdots + \sqrt{a_n} X_{\s_1\s_2\ldots \s_n}, \quad \hbox{if }\s=\s_1\s_2
 \ldots \s_n,
\Eq(cont.5)
$$  
where $a_i$ is the increment of $A$ at the step point 
$q_i=\sum_{j=1}^i \frac{\ln \a_j}{\ln 2}$, and $\s=\s_1\s_2\dots\s_n$
with $\s_i\in \{-1,1\}^{\ln \a_i N}$. 

Note that in the SK class, neither is it possible to construct such a 
represntation, nor are step functions allowed as covariances.

The representation \eqv(cont.5) allows explicit computations of 
the partition function. This was done first by Derrida and 
Gardner \cite{DG1}, 
and in full generality (and with full rigor) by Cappocaccia,
Cassandro, and Picco \cite{CaCaPi}. While we will not reproduce this 
calculations (they are in spirit not very different from those in the REM
and make use of \eqv(cont.5) to set up a recursive scheme), we will state their
result in a particularly useful form.

Let us denote the  
convex hull of the function $A(x)$ by $\bar A(x)$.
We will also need the left-derivative of this function,
$
\bar a(x)\equiv \lim_{\e\downarrow 0} \e^{-1}(\bar A(x)-\bar A(x-\e))
$
which exists for all values of $x\in (0,1]$.

\theo{\TH(CCP)} {\it Whenever $A$ is a step function with finitely many 
steps, the free energy $F_{\b,N}\equiv \frac 1N\ln Z_{\b,N}$ converges 
almost surely to the non-random limit $F_\b$ given by 
$$
F_\b=\sqrt{2\ln 2} \b\int_{0}^{x_{\b}} \sqrt{\bar a(x)} dx + 
\frac {\b^2}2 (1-\bar A(x(\b)))
\Eq(cont.6)
$$
where 
$$
x_{\b}\equiv \sup\left(x|\bar a(x)>\frac {2\ln 2}{\b^2}\right)
\Eq(cont.7)
$$
}

It is also very easy to derive from \eqv(cont.6) an explicit fromula for the
distance-distribution function
$$
f_{\b,N}(x)\equiv \mu^{\otimes 2}_{\b,N}(\dist(\s,\s')<x)
\Eq(cont.8)
$$
This just makes use of the fact that 

\proposition {\TH(GG.1.bis)}{\it For any value of $\b$, and any $i=1,\dots,n$,
$$
\E\frac{d}{d\sqrt a_i} F_{\b,N}  =-\b^2\sqrt{a_i}\E f_{\b,N}(q< q_i)
\Eq(cont.9)
$$
with the convention that $q_0=0$ and $q_n=1$.
}

This implies in fact immediately that 

\theo{\TH(OV)}{\it  Whenever $A$ is a step function with finitely many 
steps, the $f_{\b,N}$ converges in mean to the limiting function
$f_{\b}$ with
$$
\E f_\b(x)=\cases
\b^{-1}\sqrt{2\ln 2}/\sqrt{\bar a(x)},&\hbox{\rm if}\, x\leq x_\b\cr
                 1,&\hbox{if}\, x> x_\b\endcases
\Eq(cont.11)
$$
}

It is obvious that if $A_n$ is a sequence of step functions that converges to 
a limiting function $A$, then the sequences of free energies and distance 
distributions converge.  It in not very difficult to show \cite{BK3} that
these limits then are in fact the free energies and distribution functions
for the corresponding models with arbirtrary $A$. 
The results obtained here coincide with those of Derrida and Gardner, and 
in particular reproduce exactly the findings of the replica method \cite {DG2}.

\smallskip
\line{\bf \ver.2. Gibbs measures and point processes.\hfill}

As in the case of the REM, Ruelle \cite{Ru} had proposed an effective model for the 
thermodynamic limit of the GREM in terms of Poisson processes, or rather 
{\it ``Poisson cascades''}, i.e. nested sequences of Poisson processes, without
establishing a rigorous relation between the two models.  Ruelle also 
constructed limiting objects of his processes when the number of ``levels''
(i.e. $n$) tends to infinity. The connection between Ruelle's models and the 
GREMs with finitely many levels have been made rigorous in \cite{BK1}.
While again in spirit the proofs are similar to those in the REM, they 
require considerably more computations. 

However, it is quite remarkable that via the Ghirlanda-Guerra relations,
one can construct (at least in principle) the thermodynamic limit on the 
level of the measures on the mass distribution without much explicit 
computation even in the case of continuous  $A$. To prove these
inequalities, we have to impose a ``non-criticality'' conditions on
$A$:  For any $x$ where the convex hull of $A$ is not stricltly convex
(i.e. where $\bar A$ is linear in neighborhood of $x$, $A(x)<\bar
A(x)$). We assume this condition to hold in the remainder of the article.

It will be convenient to introduce here the analogues of the random measures
$\KK$ defined above where the overlap $R_N$ is replaced by the distance
$\dist$. 
I.e. we set now 
$$
m_\s(x)\equiv \mu_{\b,N}(\s:\dist(\s',\s)> x)
\Eq(C.71)
$$
and 
$$
\KK_{\b,N} \equiv\sum_{\s\in \SS_N} \mu_{\b,N}(\s)\d_{m_\s(\cdot)}
\Eq(good.1)
$$
In the case when $A$ is a step function with finitely many steps, 
one can control the convergence of $\KK_{\b,N}$ to a limit 
rather explicitely. We will present the corresponding results, without proof, 
below.

In the general case, this will no longer be possible. However, the 
 Ghirlanda-Guerra identities will allow again to prove the existence of the
limit and to decribe its properties. The key point to notice is that
to prove convergence, it is enough to prove convergence of all expressions of
the form 
$$
\eqalign{
&\E \Biggl(\left(\int \KK_{\b,N}(dm) m(\D_{11})^{r_{11}}
\dots m(\D_{1j_1})^{r_{1j_1}}\right)^{q_1}\dots\cr
&\quad\quad\dots
\left(\int \KK_{\b,N}(dm) m(\D_{l1})^{r_{l1}}
\dots m(\D_{lj_l})^{r_{lj_l}}\right)^{q_l}\Biggr)
}
\Eq(topo.1)
$$
where $\D_{ij}\subset [0,1]$ and 
$q_i,r_{ij}$ are integers.

The key is thus to establish again the Ghirlanda-Guerra identities. 
In this the process $X_\s(t)$ plays a crucial r\^ole.
It will be  convenient to use the time-changed process
$$
Y_\s(t)\equiv X_\s(A(t)) 
\Eq(conv.1)
$$

\theo{\TH(CON.11)}{\it For any $n\in \N$ and any $x\in [0,1]\ba x_\b$,
$$
\eqalign{
\lim_{N\uparrow\infty}
&\Biggl|\E\mu^{\otimes
n+1}_{\b,N}\left(h(\s^1,\dots,\s^n)\1_{A(\dist(\s^k,\s^{n+1}))\geq x}
\right)
\cr&-\frac 1n\E\mu^{\otimes n+1}_{\b,N}\left(h(\s^1,\dots,\s^n)\left(
\sum_{l\neq k}^n \1_{A(\dist(\s^k,\s^l))\geq x}+\E\mu^{\otimes 2}_{\b,N}
(\1_{A(\dist(\s^1,\s^2))\geq x}
)\right)\right)\Biggr|=0
}
\Eq(cont.22)
$$
}

\proof As a first step we need the following lemma.

\lemma{\TH(CON.12)}{\it  Let 
$h:\SS_N^n \rightarrow \R$ be any bounded
function of $n$ spins. For any $t\in (0,1] $
$$
\eqalign{
&\frac1{\sqrt N}\E\mu_{\b,N}^{\otimes
n}\left(dY_{\s^k}(t)h(\s^1,\dots,\s^n)\right)
\cr&=\b \E\mu_{\b,N}^{\otimes
n+1}\left(h(\s^1,\dots,\s^n)\left(\sum_{l=1}^n
\1_{\dist(\s^k,\s^l)\geq t} 
-n\1_{\dist(\s^k,\s^{n+1})\geq t}\right)\right)dA(t)
}
\Eq(cont.23)
$$
}

\proof The proof makes use of the Gaussian integration by parts formula
$$
\eqalign{
\E dX_\s(t) f\left(\int dX_{\s'}(s)\right)
&=\E f'\left(\int dX_{\s'}(s)\right) \int \E dX_\s(t)dX_{\s'}(s)\cr
&=\E f'\left(X_{\s'}\right) \1_{A(\dist(\s,\s'))\geq t} dt
}
\Eq(cont.24)
$$
where $f$ is any differentiable function. 
Note that the left hand side of  \eqv(cont.23)
can be written as
$$
N^{-1/2}\E \E_{\s^1\dots\s^n} h(\s^1,\dots,\s^n)
 dY_{\s^k}(t) \prod_{l=1}^nf\left(X_{\s^l}\right)
\Eq(cont.25)
$$
with 
$$
 f\left(X_{\s^l}\right)=
\frac{ e^{\b\sqrt N  X_{\s^l}(1)}} 
{\E_{\s^l} e^{\b\sqrt N  X_{\s^l}(1)}}
\Eq(cont.26)
$$
Using \eqv(cont.24) gives readily 
$$
\eqalign{
&\frac1{\sqrt N}\E\mu_{\b,N}^{\otimes
n}\left(dY_{\s^k}(t)h(\s^1,\dots,\s^n)\right)
\cr&=\b \E\mu_{\b,N}^{\otimes
n+1}\left(h(\s^1,\dots,\s^n)\left(\sum_{l=1}^n
\1_{A(\dist(\s^k,\s^l))\geq t} 
-n\1_{A(\dist(\s^k,\s^{n+1}))\geq t}\right)\right)dt
}
\Eq(cont.27)
$$
Realizing that $A(\dist(\s,\s'))< A(t)$ is equivalent to $\dist(\s,\s')< t$
whenever $A(t)$ is not constant then yields the claim of the lemma.
\endproof

The more important step of the proof is contained in the next lemma.

\lemma {\TH(CON.13)}{\it Let $h$ be as in the previous lemma.
 Except possibly when $t=x_\b$,
$$
\eqalign{
&\lim_{N\uparrow\infty}\frac 1{\sqrt N}\Biggl|
\E\mu_{\b,N}^{\otimes
n}\left((Y_{\s^k}(t)-Y_{\s^k}(t-\e))h(\s^1,\dots,\s^n)\right)\cr
&\quad-
\E\mu_{\b,N}\left( Y_{\s^k}(t)-Y_{\s^k}(t-\e)  \right)
\E\mu_{\b,N}^{\otimes
n}\left(h(\s^1,\dots,\s^n)\right)\Biggr|=0
}\Eq(cont.24.1)
$$
}

\proof Let us write 
$$
\eqalign{
&\left(
\E\mu_{\b,N}^{\otimes
n}\left(Y_{\s^k}(t)-Y_{\s^k}(t-\e)\right)-
\E\mu_{\b,N}\left(Y_{\s^k}(t)-Y_{\s^k}(t-\e)\right)
\E\mu_{\b,N}^{\otimes
n}\left(h(\s^1,\dots,\s^n)\right)\right)^2\cr
&=\left(
\E\mu_{\b,N}^{\otimes
n}\left(\bigl((Y_{\s^k}(t)-Y_{\s^k}(t-\e))-\E\mu_{\b,N}^{\otimes
n}(Y_{\s^k}(t)-Y_{\s^k}(t-\e))\bigr) h(\s^1,\dots,\s^n)\right)\right)^2\cr
&\leq 
\E\mu_{\b,N}^{\otimes n}\left((Y_{\s^k}(t)-Y_{\s^k}(t-\e))-
\E\mu_{\b,N}^{\otimes
n}(Y_{\s^k}(t)-Y_{\s^k}(t-\e))\right)^2 \E\mu_{\b,N}^{\otimes n}
\left( h(\s^1,\dots,\s^n)\right)^2
}
\Eq(cont.30)
$$
where the last inequality is the Cauchy--Schwarz inequality applied to the 
joint expectation with respect to the Gibbs measure and the disorder.
Obviously the first factor in the last line is equal to 
$$
\eqalign{
&\E\mu_{\b,N}\left((Y_{\s^k}(t)-Y_{\s^k}(t-\e))-
\mu_{\b,N}(Y_{\s^k}(t)-Y_{\s^k}(t-\e))\right)^2
    \cr
&\quad+\E\left( \mu_{\b,N}(Y_{\s^k}(t)-Y_{\s^k}(t-\e))-\E
\mu_{\b,N}(Y_{\s^k}(t)-Y_{\s^k}(t-\e))\right)^2 
}
\Eq(cont.31)
$$
Now let us introduce the deformed process
$$
X^u_\s \equiv X_\s +u\left(Y_\s(t)-Y_\s(t-\e)\right)
\Eq(cont.32)
$$
If we denote by $F^u_{\b,N}$ the free energy 
corresponding to this deformed 
process, the last line of \eqv(cont.31)
can be represented as 
$$
 \b^{-2}   \E \frac {d^2}{du^2} F^u_{\b,N}
+N
\b^{-2}\E\left( \frac {d}{du} F^u_{\b,N} -\E\frac
{d}{d u} F^u_{\b,N} \right)^2 
\Eq(cont.33)
$$

At this point we need a concentration result on the free energy which we state
here without proof.

\lemma {\TH(CONC.0)}{\it For any $\b$, and any covariance distribution 
$A$, for any $\e\geq 0$
$$
\P\left[\left|F_{\b,N}-\E F_{\b,N}\right|>r\right]
\leq 2\exp\left(-\frac {r^2N}{2\b^2 }\right)
\Eq(conc.0)
$$
}

 $ F^u_{\b,N}$ converges as $N\uparrow\infty$ and that the 
limit is infinitely differentiable as a function of $u$,
 except possibly when $x_\b=t$, provided $A$ satisfies the
 non-criticality condition; moreover, $-F^u_{\b,N}$ is convex in
the variable  $u$. This can be seen by explicit computation using the 
expression \eqv(cont.6) for the free energy. 
 Then a
standard result of convex analysis (see \cite {Ro}, Theorem 25.7) imply that 
$$
\limsup_{N\uparrow\infty} (- \E \frac {d^2}{du^2} F^u_{\b,N})
=- \frac {d^2}{d u^2}   \lim_{N\uparrow \infty}\E  F^u_{\b,N}
\Eq(C.59bis)
$$
which is finite at zero except possibly if $x_\b=t$.
 Thus, the first term in 
\eqv(C.58)  will vanish
 when divided by $N$. To see that the coefficient of $N$
of the second term gives a vanishing contribution, we use the general fact that
if the variance of family  of a convex (or concave) functions tends to 
zero, then the same is true for its derivative, provided the second
derivative
of the expectation is bounded (see e.g. Lemma 8.9 in \cite{BG},
or Proposition 4.3 in \cite{T2}).

But by Lemma \thv(CONC.0)
 the variance of $F_{\b,N}$ tends to zero, and    \eqv(C.59bis)
implies that  $\E \frac {d^2}{du^2} F^u_{\b,N}$ is bounded for large
 enough $N$ whenever $ \frac {d^2}{d u^2} \E  F^u_{\b}$
is finite.
 Hence the result of the 
lemma is proven. \endproof

To prove the theorem we use integrate \eqv(cont.23) and then use 
\eqv(cont.24.1) on the left hand side.
This gives 
$$
\eqalign{
&\lim_{N\uparrow\infty}
\Biggl(\frac1{\sqrt N}\E\mu_{\b,N}^{\otimes
n}\left(Y_{\s^k}(t)-Y_{s^k}(t-\e)\right)
\E\mu_{\b,N}^{\otimes n}
\left(h(\s^1,\dots,\s^n)\right)\cr
&-\b\int_{t-\e}^t\left( \E\mu_{\b,N}^{\otimes
n+1}\left(h(\s^1,\dots,\s^n)\left(\sum_{l=1}^n
\1_{\dist(\s^k,\s^l)\geq s} 
-n\1_{\dist(\s^k,\s^{n+1})\geq s}\right)\right)\right)dA(s)\Biggr)=0
}
\Eq(cont.27.1)
$$
Finally, we use once more \eqv(cont.23) with $n=1$ to express 
$\E\mu_{\b,N}^{\otimes
n}\left(Y_{\s^k}(t)-Y_{s^k}(t-\e)\right)$ in terms of the two replica
distribution. The final result follows by trivial algebraic manipulations
and the fact that $\e$ is arbitrary.
\endproof\endproof

Following [GG], we  now define  the family of measures   $\Q^{(n)}_N$ 
on the space $[0,1]^{n(n-1)/2}$. 
$$
\Q_{\b,N}^{(n)}(\un \dist\in \AA)\equiv \E\mu_{N,\b}^{\otimes n} 
\left[\un \dist \in \AA\right]
\Eq(10.6.1)
$$
where $\un \dist $ denotes the vector of replica distances 
whose components are
$\dist(\s^l,\s^k)$, $1\leq l<k\leq n$. Denote by $\BB_k$ the sigma-algebra 
generated 
 by the first $k(k-1)/2$ coordinates, and let
 $A$ be a Borel set in $[0,1]$. 

\theo{\TH(GGG)}{\it The family of measures $\Q_{\b,N}^{(n)}$ converge
to limiting measures $\Q_{\b}^{(n)}$ for all finite $n$, as $N\uparrow\infty$.
Moreover, these measures are uniquely determined by the distance distribution 
functions $f_\b$. They satisfy the identities
$$
\Q^{(n+1)}_{\b} \left(d_{k,n+1}\in A  | \BB_{n}\right)
=\frac 1n \Q_\b^{(2)}(A) +\frac 1n\sum_{l\neq k}^n
\Q^{(n)}_{\b} \left(d_{k,l}\in A  | \BB_{n}\right)
\Eq(GGG.1)
$$
for any Borel set $A$.
}

\proof
Choosing  $h$ as the indicator function of 
any desired event in $\BB_k$, one sees that \eqv(cont.22) implies  \eqv(GGG.1).
This  actually implies that in the limit $N\uparrow\infty$,
the  family of measures $\Q^{(n)}_{\b,N}$ is entirely
determined by the two-replica distribution function. While this 
may not appear 
obvious,  it follows when taking into account the ultrametric property of the 
function $\dist$. This is most easily seen by realising that the prescription 
of the mutual distances between $k$ spin configurations amounts to prescribing 
a tree  (start all $k$ configurations at the origin and continue 
on top of each other as long as the coordinates coincide, then branch of).
To determine the full tree of $k+1$ configurations, it is sufficient to 
know the overlap of configuration $\s^{(k+1)} $ with the configuration it
has maximal overlap with, since then all overlaps with all other configurations
are determined. But the corresponding probabilities can be computed
recursively via \eqv(GGG.1).

\centerline{\epsffile{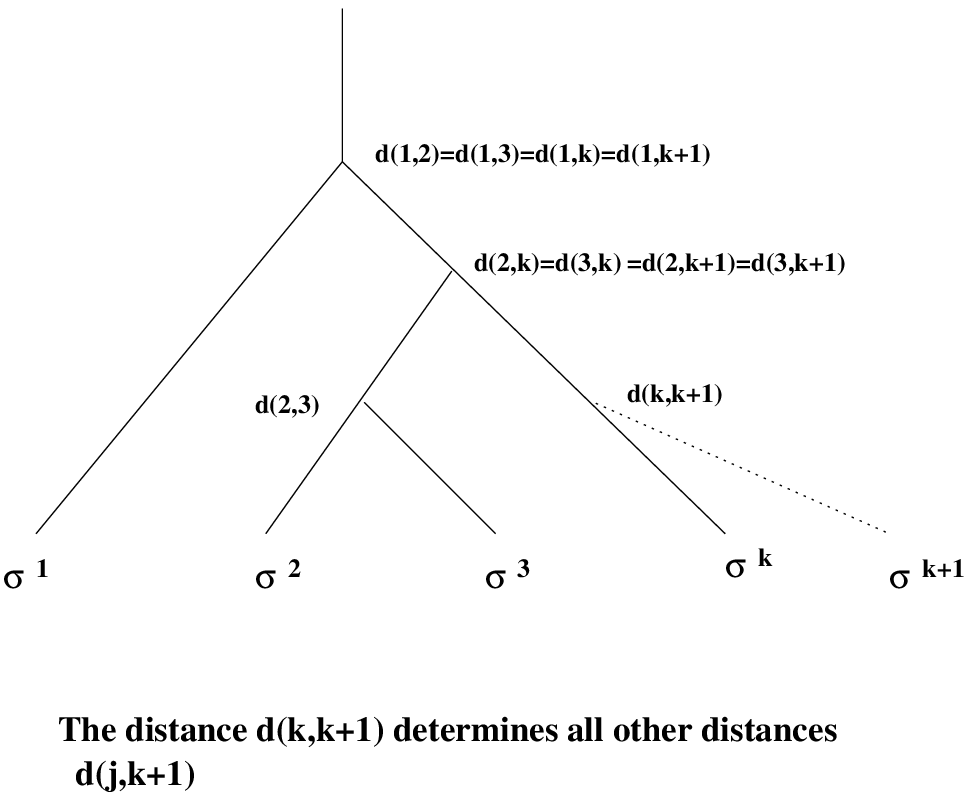}}

Now we have already seen that  $\Q_{\b,N}^{(2)}=
\E \tilde f_{\b,N}$ converges. Therefore the relation \eqv(GGG.1) implies the
convergence
of all distributions $\Q^{(n)}_{\b,N}$, and proves the relation 
\eqv(GGG.1) hold for the limiting measures. \endproof

Now it is clear that all expressions of the form \eqv(mass.1) (with 
$R_N$ replaced by $\dist$)
can be expressed in terms of the measures $\Q_{\b,N}^{(k)}$
for $k$ sufficiently large (we leave this as an exercise for the 
reader to write down). Thus, Theorem \thv(GGG) implies in turn the 
convergence of the process $\KK_{\b,N}$ to a limit $\KK_\b$.

A remarkable feature takes place again if we are only interested in the 
marginal process $K_\b(t)$ for fixed $t$. This process is a simple point 
process on $[0,1]$ and is fully determined in terms of the moments 
$$
\eqalign{
&\E\left(\int K_{\b,N}(t)(dx) x^{r_1} \dots\int  K_{\b,N}(t)(dx) x^{r_j}\right)
\cr
&=\E \mu_{\b,N}^{\otimes r_1+\dots +r_j+j}
\Bigl(\dist(\s^1,\s^{j+1})>t,
\dots,\dist(\s^1,\s^{j+r_1})>t,\dots,\cr
&\quad\dots,
\dist(\s^j,\s^{j+r_1+\dots+r_{j-1}+1})>t,\dots
 \dist(\s^j,\s^{j+r_1+\dots+r_{j}})>t\Bigr)
}\Eq(good.20)
$$
This restricted family of moments satisfies via the Ghirlanda-Guerra identities
exactly the same recursion as in the case of the REM. This implies:

\theo{\TH(GGG.2)}{\it Assume that $t$ is such that $\E\mu^{\otimes 2}_\b
(\dist(\s,\s')<t)=1/\a>0$. Then the random measure $K_\b(t)$ is a 
Dirichlet-Poisson process (see e.g. \cite{Ru,T1}) with parameter $\a$ .}

In fact much more is true. We can consider the processes on arbitrary finite 
dimensional marginals, i.e.
$$
K_{\b,N}(t_1,\dots,t_m)\equiv \sum_{\s\in \SS_N}
\mu_{\b,N}(\s) \d_{m_\s(t_1),\dots,m_{\s}(t_m)}
\Eq(good.21)
$$
for $0<t_1<\dots<t_m<1$. The point is that this process is 
entirely determined by the expressions \eqv(topo.1)
with the $\D_{ij}$ all of the form $(t_i,1]$ for $t_i$ in the fixed set 
of values $t_1,\dots,t_m$. This in turn implies that the process is determined
by the multi-replica distribution functions $\Q_{\b,N}^{(n)}$ restricted to 
the discrete set of events $\{\dist(\s^i,\s^j)>t_k\}$. 
Since these numbers are totally determined through the Ghirlanda-Guerra 
identities, they are identically to those obtained in a GREM with $m$
levels, i.e. a function $A$ having steps at the values $t_i$, those 
two-replica distribution function takes the same values 
as  that of the model with continuous $A$ at the points $t_i$ and is 
constant between those values. 
In fact

\theo{\TH(GGG.3)}{\it Let $0<t_1<\dots<t_k\leq q_{max}(\b)$ be
points of increase of $\E f_\b$.    Consider
 a GREM with $k$ levels and parameters
$\a_i,a_i$ and temperature $\tilde \b$ that satisfy
$\ln\a_i/\ln 2=t_i-t_{i-1}$, $\tilde\b^{-1}\sqrt{2\ln \a_i/a_i}=\E
f_\b(t_i)$.
Then 
$$
\lim_{N\uparrow\infty}\KK_{\b,N}(t_1,\dots,t_k)=
\KK^{(k)}_{\tilde \b }
\Eq(good.22)
$$
}

Thus, if the $t_i$ are chosen in such a way that for all of them $
\E f_\b(t_i)>0$, then we can construct an explicit representation of the 
limiting marginal process $\KK_\b(t_1,\dots,t_m)$ in terms of a Poisson-cascade
process via the corresponding formulae in the associated $m$-level
GREM.  This construction is done in the next section.
 In this sense we obtain an explicit description of the
limiting mass distribution function $\KK_\b$.

\line{\bf \ver.2. Probability cascades in the GREM with finitely many levels.\hfill}

Let us now briefly explain the structure of the process $\KK_\b$ in the 
case when $A_n$ is a step function
with steps of hight $a_i$ at the values $t_i\equiv \frac {\ln \a_i}{\ln 2}$.
To avoid complications, we will assume that
the linear interpolation of this function is convex, and that all 
points $t_i$ belong to the extremal set of the convex hull.

\remark
I will not give the proofs here, that are somewhat involved,   in particular
when the general case is considered. They can be found in \cite{BK1, BK2}.
The following summary of results is in fact just a cooked down version 
of the complete analysis of the GREM with finitely many hierarchies 
given there. Note that we draw heavily on the representation \eqv(cont.5).

We introduce the function $u_{\ln \alpha, N}(x)$, $x\in \R$ as
$$  
u_{\ln \a, N}(x)=\sqrt{2\ln \alpha N}+\frac{x}{\sqrt{2 \ln \alpha N}} 
  -\frac{ \ln N +\ln \ln  \alpha + \ln 4\pi }{2
 \sqrt {2 \ln \alpha N}}.
\Eq(an.3.1)
$$   
Note that then for all $i$,
$$
\sum_{\s_i}\d_{u_{\ln \a_i,N}^{-1}(X_{\s_1\dots\s_{i-1}\s_i})}
\rightarrow \PP_i
\Eq(an.3.2)
$$
where $\PP_i$ are all independent Poisson point processes on $\R$ with 
intensity measure $e^{-x}dx$. Then under the assumptions on $A$, the following 
result holds:

\theo{\TH(1.2bis)} {\it   The  following point processes on $\R^k$
 $$
\PP_N^{(k)}\equiv \sum_{\s_1}\delta_{u^{-1}_{\ln\a_1, N}(Y_{\s_1})}  
     \sum_{\s_2}\delta_{u^{-1}_{\ln \a_2,N}(Y_{\s_1\s_2})}  
   \cdots 
   \sum_{\s_k}\delta_{u^{-1}_{\ln\a_k ,N}(Y_{\s_1\s_2\ldots \s_k})}  
   \rightarrow  \PP^{(k)}
$$
converge weakly to point process $\PP^{(k)}$ on $\R^k$, 
   which is characterised by the following generating 
   functions:
$$
\eqalign{
   F_{\Delta_1\times \cdots \times \Delta_k}(z) &
   \equiv \E z^{\sum_{x_1}\1_{\{ x_1\in \Delta_1 \}}\cdots
    \sum_{x_k}\1_{\{x_k\in \Delta_k \}}} \cr
 & =f_{1,\Delta_1}( f_{2, \Delta_2}(f_{3, \Delta_3} \cdots 
    ( f_{k-1, \Delta_{k-1}} ( f_{k, \Delta_k}
    (z)))\cdots )),\qquad |z|<1  } 
\Eq(gf)
$$
where $f_{i, \Delta_i}(z)=e^{K
_i(z-1)(e^{-a_i}- e^{-b_i})}$, 
   $\Delta_i=(a_i, b_i]$ with $a_i, b_i \in  \R$ or $b_i=\infty$,
  $i=1,2,\ldots, k$. 

\noindent    Moreover,  the following  independence 
     properties  of  the counting random variables of the process 
$\PP^{(k)}$,
    $\sum_{x_1}\1_{\{x_1\in \Delta^j_1 \}}\cdots 
    \sum_{x_k}\1_{\{x_k\in \Delta^j_k \}}$, corresponding  
   to the  intervals 
    $\Delta_1^j \times\cdots \times  \Delta_k^j$, $\Delta_i^j=[a_i^j, b_i^j)$, 
   $j=1,2,\ldots, k$, $k>1$,  hold true:
           
\noindent (i) If the first components of these intervals are disjoint, 
     i.e.\ $a_1^1\leq b_1^1\leq a_1^2\leq b_1^2 \leq 
     \cdots a_1^k \leq b_1^k$, then these r.v. are independent.

\noindent (ii) If the first $l-1$ components of these intervals  coincide 
      and the  $lth$ components  are disjoint, i.e.
     $\Delta_i^1=\cdots =\Delta_i^k$ for $i=1,\ldots,l-1$ 
     and $a_l^1\leq b_l^1\leq a_l^2\leq b_l^2 \leq 
     \cdots a_l^k \leq b_l^k$, then  these r.v.\ are  conditionally 
      independent under condition that  
     $\sum_{x_1}\1_{\{x_1\in \Delta_1\}}\cdots 
      \sum_{x_{l-1}}\1_{\{x_{l-1}\in \Delta_{l-1}\}}$ is fixed. }

\remark  This theorem was proven for $k=2$  in \cite{GMP}.

      We would like to clarify an intuitive construction 
      of  the process $\PP$.   If $k=1$, this is          
      just a Poisson point process on $\R$ 
      with  intensity measure $K_1e^{-x}dx$.
      To construct $\PP$  on $\R^2$ for $k=2$ we place 
      the process $\PP$ for $k=1$ on the axis of the first coordinate  
      and through each of its points 
      draw a straight line parallel to the axis 
      of the second coordinate. Then we put on each of 
      these lines independently 
      a Poisson point process with intensity 
      measure $K_2 e^{-x}dx$. These points on $\R^2$ form
      the process  $\PP$ 
      with $k=2$.  Whenever $\PP$ is constructed 
      for $k-1$, we place it on the plane of the first 
      $k-1$ coordinates and through each of its points 
      draw a straight line parallel to the axis of the 
      $n$th coordinate. On each of these lines we put
      after  independently a Poisson point process 
      with intensity measure $K_k e^{-x}dx$. These points 
      constitute  $\PP$ on $\R^k$. Indeed, the projection of $\PP^{(k)}$ 
      in $\R^k$ to the plane of the first $\ell$ coordinates 
      is distributed as the process $\PP^{(\ell)}$ in $\R^\ell$.

We are now also in the position to formulate a result on the extreme
order statistics of the random variables $X_\s$.

Let  $\gamma_l\equiv  \sqrt{ a_l}/\sqrt{2\ln  \a_l}$,
   $l=1,2, \ldots, n$. 
   By our assumption on $A$,   $\gamma_1>\gamma_2>\cdots>\gamma_n$.   
 Define the function $U_{J,N}$ by
$$
U_{J,N}(x)\equiv \sum_{l=1}^n \left(
\sqrt{2N a_l\ln \bar\a_l}-N^{-1/2} \g_l(\ln( N (\ln \a_l))+\ln 4\pi)/2
\right)+N^{-1/2}
x
\Eq(ex.1)
$$
and the point process
$$
\EE_N \equiv \sum_{\s\in \{-1,1\}^N} \d_{U_{J,N}^{-1}(X_\s)}.
\Eq(ex.2)
$$
Then the following holds true:

\theo{\TH(ex1.5)}{\it  The point process $\EE_N$ converges weakly, as
$N\uparrow\infty$, to the 
point process on $\R$
$$
\EE\equiv \int_{\R^n} \PP^{(n)}(dx_1,\dots,dx_n) \d_{\sum_{l=1}^m \g_lx_l}
\Eq(ex.3)
$$
where $\PP^{(n)}$ is the Poisson cascade introduced in Theorem \thv(1.2bis).
 }
   

Next we state a convergence result for the partition function that is 
analogous to the low-temperature result Theorem \thv(4.2), (v), in the REM.
 
One would be tempted to believe that the process that is relevant for
the extremal process will again be the right one to choose. However,
this will be the case only for large enough $\b$. 
  However,   only the first $l(\b)$ levels of the process participate, where
$$ 
l(\beta)\equiv \max\{l\geq 1:  \beta^2\g_l>1\} 
\Eq(lb)
$$     
  and $l(\beta)\equiv 0$ if $\beta^2 \g_l \leq 1$.

      The following theorem yields  the fluctuations 
 of the partition function and connects the GREM to Ruelle's processes.

\theo{\TH(1.5bis)}{\it  
 With the definitions above, under our hypothesis on $A$,
$$
\eqalign{
& e^{ \sum_{j=1}^{l(\beta)} \big(
        -\beta N \sqrt{2 a_j \ln  \a_j}+ 
       \beta  \gamma_j [\ln (N\ln  \a_j) +\ln 4\pi]/2 + N \ln \a_j\big)
    - N \sum_{i={l(\beta)}+1}^n \beta^2 a_i/2 }Z_{\beta, N} \cr
   & \limlaw  C(\beta) \int_{\R^{l(\beta)}  } 
     e^{\beta \gamma_1 x_1+ \beta \gamma_2 x_2+\cdots+ \beta \gamma_{l(\beta)}
       x_{l(\beta)}} \PP^{(l(\b))}(dx_1\ldots dx_{l(\b)} ). 
}
\Eq(mainc)
$$
   This integral is 
  over the process $\PP^{(l(\b))}$ on $\R^{l(\b)}$ 
   constructed in Theorem \thv(1.2bis) .
  The constant $C(\b)$ satisfies
$$
C(\beta)=1,
\text{if} \beta \gamma_{l(\b)+1}<1,
\Eq(cb1)
$$ 
  and 
$$
C(\b)= P\Big(\bigcap_{i:   {l(\b)}+1 \leq i \leq {l(\b)+1} \atop
                  ( a_{{l(\b)}+1}+\cdots +a_i)/  a_{{l(\b)+1}}
                    = \ln (\a_{{l(\b)}+1} \cdots \a_i)/\ln \bar
                   \a_{{l(\b)+1}}    } 
\kern -1em (\sqrt{a_{{l(\b)}+1}}Z_{{l(\b)}+1}+\cdots + \sqrt{a_i} Z_i<0) \Big) 
\Eq(cb2)
$$
  if $\beta \gamma_{l(\b)+1}=1$\hfill\break
 where $Z_{{l(\b)}+1}, \ldots, 
  Z_{{l(\b)+1}} $ are independent standard Gaussian r.v. 
   Moreover  }
$$ 
\eqalign{
 \ln Z_{N,\beta} -  \E \ln Z_{N,\beta} 
 &   \limlaw   
    \ln C(\beta) \int_{\R^{l(\beta)}  } 
     e^{\beta \gamma_1 x_1+ \beta \gamma_2 x_2+\cdots+ \beta \gamma_{l(\beta)}
       x_{l(\beta)}} \PP(dx_1\ldots dx_{l(\b)}).} $$

Let us introduce the sets 
$$
B_l(\s)\equiv \{\s'\in\SS_N:\dist(\s,\s')\geq q_l\}
\Eq(pp.1)
$$
We define point processes $\WW_{\b,N}^m$
on $(0,1]^m$ given by
$$
\WW_{\b,N}^m\equiv 
\sum_{\s} \d_{\left(\mu_{\b,N} \left(B_1(\s)\right),\dots,
\mu_{\b,N}\left(B_m(\s)\right)\right)}\frac {\mu_{\b,N}(\s)}
{\mu_{\b,N}\left(B_m(\s)\right)}
\Eq(gi.1)
$$
as well as their projection on the last coordinate,
$$
\RR_{\b,N}^m\equiv 
\sum_{\s} \d_{
\mu_{\b,N}\left(B_m(\s)\right)}\frac {\mu_{\b,N}(\s)}
{\mu_{\b,N}\left(B_m(\s)\right)}
\Eq(pp.2)
$$
It is easy to see that the processes $\WW_{\b,N}^m$ satisfy 
$$
 \WW_{\b,N}^m(dw_1,\dots,dw_m) =\int_{0}^1 W^{m+1}_{\b,N}(dw_1,\dots, 
dw_m,d{w_{m+1}} )\frac {w_{m+1}}{w_m}
\Eq(pp.3)
$$
where the  integration is of course over the last coordinate $w_{m+1}$.
Note that these processes will in general not all converge, but will do so 
only when for some $\s$, $\mu_{\b}(B_m(\s))$ is strictly positive. From 
our experience with the partition function, 
it is clear that this will be the case
precisely when $m\leq l(\b)$. In fact, we will prove that

\theo{\TH(GI.1)}{\it If $m\leq l(\b)$, the point process $\WW_{\b,N}^m$ on
 $(0,1]^{m}$ 
converges 
weakly to the point process $\WW^m_\b$ whose atoms $w(i)$ 
are given in terms of the 
atoms $(x_1(i),\dots,x_m(i))$ of the point process $\PP^{(m)}$ by
$$
\eqalign{
&(w_1(i),\dots,w_{m}(i))\cr
&= \left(\frac {\int \PP^{(m)}(dy)\d(y_1-x_1(i)) 
e^{\b(\g,y)}}{
\int \PP^{(m)}(dy)e^{\b(\g,y)}},\dots,\frac{\int \PP^{(m)}(dy)\d(y_1-x_1(i))\dots
\d(y_{m}-x_{m}(i)) e^{\b(\g,y)}}{
\int \PP^{(m)}(dy)e^{\b(\g,y)}}\right)
}
\Eq(gi.2)
$$
and the point processes $\RR^{(m)}_{\b,N}$ converge to the 
point process $\RR^{(m)}_\b$ 
whose atoms are the last component of the atoms in 
\eqv(gi.2).
}

Of course the most complete object we can reasonably study is the 
process $\wh \WW_\b\equiv \WW_\b^{l(\b)}$. Analogously, we will set
 $\wh\RR_\b\equiv \RR_\b^{l(\b)}$.

The point processes $\wh\WW^{(m)}_\b$ takes values on vectors 
whose components form increasing sequences in $(0,1]$. Moreover,
these atoms are naturally clustered in a hierarchical way. These processes 
were introduced by Ruelle \cite{Ru} and called {\it probability
  cascades}.   Finally, our last  theorem gives the explicit construction 
  of the limiting process $\KK_{\beta}$ in the case of the step-
  function $A$ via Ruelle's probability cascades.

\theo{\TH(GI.K)}{\it 
 The process $\KK_{\b,N}$ converges 
  to the process $\KK_{\beta}$ which is  supported  on measures 
   $\delta_{w}$  indexed by points 
  $w=(w(1),\ldots, w(l(\beta))) \in \wh \WW_{\beta}$. More
   precisely 
   $$
   \KK_{\beta}=\int_{\R^{l(\b)}}\wh\WW_{\beta}(dw)w(l(\beta))
\delta_{m(w)}.$$
where the measure $m(w)$ is given by the formula
$$
m(w)= (1-w_1)\d_{0}+(w_1-w_2)\d_{\ln \a_1/\ln 2}+\dots+w_{l(\b)}
\d_{\ln(\a_1\cdots \a_{l(\b)})/\ln 2}
$$
}  
\bigskip
\frenchspacing
\line{\bf References\hfill}
\medskip
{\baselineskip=8pt\parskip=4pt
\refer
\item{[AC]} M. Aizenman, P. Contucci. On the stability of the quenched
state in mean field spin glass models. J. Stat. Phys. 92, 765-783 (1998).
\item{[ALR]} M. Aizenman, J.L. Lebowitz,  D. Ruelle 
               Some rigorous results on Sherrington-Kirkpatrick
    spin glass model. { Commun. Math. Phys.} { 112} (1987),
 3-20. 
\item{[BeLe]} J. Bertoin,  J.-F. Le Gall. The Bolthausen-Sznitman coalescent 
and the genealogy of continuous-state branching
processes. Probab. Theory Related Fields { 117}, 249-266 (2000).
\item{[BoSz]} E. Bolthausen and A.-S. Sznitman. 
On Ruelle's probability cascades and an abstract cavity method.
Comm. Math. Phys. { 197} (1998), 247--276. 
\item{[B]} A. Bovier. Statistical mechanics of disordered systems. MaPhySto
Lecture Notes 10. Aarhus, 2002.
\item{[BG]}  A. Bovier, V. Gayrard. The Hopfield model as a 
generalized random mean field model. In Mathematics of spin glasses and 
neural  networks, A. Bovier, P. Picco. Eds., Progress in Probablity,
Birkh\"auser, Boston, (1997).
\item{[BKL]} A. Bovier, I. Kurkova,  M. L\"owe. The fluctuations of the free
energy in the REM and the $p$-spin SK models,
Ann. Probab. {\bf 30}, 605-651 (2002). 
\item{[BK1]} A. Bovier and I. Kurkova. Derrida's Generalized Random
 Energy models 1: 
Poisson cascades and extremal processes. preprint Universit\'e Paris
6 (2002).
\item{[BK2]} A. Bovier and I. Kurkova. Derrida's Generalized Random
Energy models 2:   Gibbs measures and probability cascades.  preprint
Universit\'e Paris 
6 (2002).
\item{[BK3]} A. Bovier and I. Kurkova. Derrida's Generalized Random
Energy models  3:   Models with continuous hierarchies.  preprint
Universit\'e Paris  6 (2002).
\item{[CN]} F. Comets,  J. Neveu. The Sherrington-Kirkpatrick
  Model of Spin Glasses and Stochastic Calculus: The High Temperature
  Case. { Commun Math. Phys.} { 166} (1995), 549-564.
\item{[D1]} B. Derrida. Random energy model: limit of a family of disordered
models, { Phys. Rev. Letts.} { 45}(1980), 79-82.
\item{[D2]} B. Derrida. Random energy model: An exactly solvable model of 
disordered systems, { Phys. Rev. B} { 24}, 2613-2626 (1981)
\item{[DW]} T.C. Dorlas, J.R. Wedagedera. Large deviations and the random 
energy model. Int. J. Mod. Phys. B 15 (2001), 1-15.
\item{[Ei]} Th. Eisele.  On a third-order phase transition. 
Comm. Math. Phys. 90 (1983),  125-159.
\item{[GMP]} A. Galvez, S. Martinez,  P. Picco. Fluctuations in 
Derrida's random energy and generalized random energy models, { J. Stat. 
Phys.}{ 54}  (1989), 515-529. 
\item{[GG]} S. Ghirlanda,  F. Guerra. { General properties of the overlap
           probability distributions in disordered spin systems.Towards Parisi
           ultrametricity}, J. Phys. A { 31}(1998) , 9144-9155.
\item{[GM]} D.J. Gross, M. M\'ezard. The simplest spin glass. Nucl. Phys.
{B 240} (1984), 431-452.
\item{[G]} F. Guerra. Broken Replica Symmetry Bounds in the Mean Field Spin Glass Model. preprint {\tt cond-mat/0205123} (2002).
\item{[GT]} F. Guerra, F.L. Toninelli. The thermodynamic limit in mean field
spin glass models. preprint {\tt cond-mat/0204280}.
\item{[K]} J.F.C. Kingman. Poisson processes. 
Oxford Studies in Probability, 3. Oxford Science Publications. 
The Clarendon Press, Oxford University
Press, New York, 1993. 
\item{[LLR]} M.R. Leadbetter, G. Lindgren, H. Rootz\'en.
  { Extremes and Related Properties of Random Sequences
    and Processes},
  Springer, Berlin-Heidelberg-New York, 1983.
\item{[L]} M. Ledoux. On the distribution of overlaps in the 
Sherrington-Kirkpatrick spin glass model. J. Statist. Phys. 100 
(2000), 871-892. 
 \item{[LT]} M. Ledoux,  M. Talagrand. Probability in Banach spaces.
 Springer, Berlin-Heidelberg-New York, 1991.
\item{[MPV]} M. M\'ezard, G. Parisi,  M.A. Virasoro. Spin-glass theory
and beyond. { World Scientific}, Singapore (1988).
\item{[NS]} C.M. Newman, D.L. Stein. Thermodynamic chaos and the structure of
short-range spin glasses. Mathematical 
aspects of spin glasses and neural networks, 243-287,
Progr. Probab., 41, Birkh\"auser Boston, Boston, MA, 1998.
\item{[OP]} E.Olivieri, P. Picco. On the existence of thermodynamics for 
the random energy model. Comm. Math. Phys. 96 (1984),  125-144.
\item{[Ru]} D. Ruelle.
A mathematical reformulation of Derrida's REM and GREM. 
{  Math. Phys} {108} (1987), 225-239 . 
\item{[SK]} D. Sherrington,  S. Kirkpatrick. Solvable model of a
spin glass. { Phys. Rev. Lett.}
{ 35} (1972), 1792-1796.
\item{[T1]} M. Talagrand. The Sherrington-Kirkpatrick model: 
A challenge for 
mathematicians. Probab. Theory Related Fields 110 (1998),  109-176. 
\item{[T2]} M. Talagrand. Rigorous results for the Hopfield model with many 
patterns. Probab. Theory Related Fields 110 (1998), 177-276. 
\item{[T3]} M. Talagrand.
Rigorous low-temperature results for the mean field $p$-spins interaction 
model. Probab. Theory Related Fields 117 (2000),
303-360.
\item{[T4]}  M. Talagrand. On the $p$-spin interaction model at low 
temperature. C. R. Acad. Sci. Paris S\'er. I Math. 331 (2000), 939-942. 
\item{[T5]}  M. Talagrand. Mean field models for spin glasses: a first course.
 Course given in Saint Flour (2000).
\item{[T6]}  M. Talagrand. Self organization in the low temperature region of a spin glass model. Preprint (2002).

}
\end